\documentclass[final]{elsarticle}

\usepackage{lineno,hyperref}
\modulolinenumbers[5]

\journal{Proceedings of the Combustion Institute}

\newcommand{\ud}{\mathrm{d}}
\newcommand{\pdiff}[2]{\frac{\partial{#1}}{\partial{#2}}}

\newcommand{\tdiff}[2]{\frac{\ud{#1}}{\ud{#2}}}

\newcommand{\vect}[1]{\mathbf{#1}}

\usepackage{xcolor}

\newcommand{\revis}[1]{{\textcolor{black}{#1}}}

\usepackage{amsmath}
\usepackage{amsfonts}
\usepackage{amssymb}

\usepackage{placeins}

\begin{document}

\begin{frontmatter}

\title{A data-driven kinematic model of a ducted premixed flame}

\author[mymainaddress]{Hans Yu}
\author[mymainaddress]{Matthew P.\ Juniper}
\author[mymainaddress,mysecondaryaddress]{Luca Magri\corref{mycorrespondingauthor}}
\ead{lm547@cam.ac.uk}
\cortext[mycorrespondingauthor]{Corresponding author}

\address[mymainaddress]{Department of Engineering, University of Cambridge, Trumpington Street, Cambridge CB2 1PZ, UK}
\address[mysecondaryaddress]{Institute for Advanced Study, Technical University of Munich, Lichtenbergstrasse 2a, 85748 Garching, Germany (visiting fellowship)}

\begin{abstract}
Reduced-order models of flame dynamics can be used to predict and mitigate the emergence of thermoacoustic oscillations in the design of gas turbine and rocket engines.
This process is hindered by the fact that these models, although often qualitatively correct, are not usually quantitatively accurate.
As automated experiments and numerical simulations produce ever-increasing quantities of data, the question arises as to how this data can be assimilated into physics-informed reduced-order models in order to render these models quantitatively accurate.
In this study, we develop and test a physics-based reduced-order model of a ducted premixed flame in which the model parameters are learned from high speed videos of the flame.
The experimental data is assimilated into a level-set solver using an ensemble Kalman filter.
This leads to an optimally calibrated reduced-order model with quantified uncertainties, which accurately reproduces elaborate nonlinear features such as cusp formation and pinch-off.
The reduced-order model continues to match the experiments after assimilation has been switched off.
Further, the parameters of the model, which are extracted automatically, are shown to match the first order behavior expected on physical grounds.
This study shows how reduced-order models can be updated rapidly whenever new experimental or numerical data becomes available, without the data itself having to be stored.
\end{abstract}

\begin{keyword}
premixed combustion\sep thermoacoustics\sep reduced-order modeling\sep data assimilation\sep uncertainty quantification
\end{keyword}

\end{frontmatter}


\clearpage


\section{Introduction}
\label{sec:intro}

Thermoacoustic oscillations are a persistent challenge in the design of jet and rocket engines~\cite{Lieuwen2005, Culick2006}.
On the one hand the {\it qualitative} mechanism is well-understood:
acoustic perturbations cause heat release rate perturbations at the flame through a variety of mechanisms~\cite{Lieuwen2005}.
These heat release rate perturbations, if sufficiently in phase with the acoustic pressure, amplify the acoustic perturbations, closing the feedback loop~\cite{Strutt1878}.
On the other hand, the {\it quantitative} prediction of thermoacoustic dynamics in real engines is difficult and computationally demanding, despite advances in large-eddy simulations~(LES) of reacting flows~\cite{Poinsot2017}.
The quantitative prediction is difficult because thermoacoustic systems are extremely sensitive to small changes to the system~\cite{Juniper2018, Magri2019}.

Recent advances in data-driven methods and machine learning introduce new approaches to the development of predictive methods~\cite{Duraisamy2019, Brunton2020}.
Most data-driven approaches identify systems and extract models using projections or kernels to map from high-fidelity data to reduced-order models~\cite{Benner2015}.
Examples include proper orthogonal decomposition, dynamic mode decomposition, active subspaces, resolvent analysis and variational autoencoders.
In this study, a different approach is developed and tested; this is a data-driven, physics-informed reduced-order model.
We begin with the $G$-equation, which is a reduced-order model of the flame surface~\cite{Dowling1999}.
The surface propagates into the fresh gas, whose velocity is prescribed.
The propagation speed is determined by fuel composition and local curvature.
Unlike projection- and kernel-based methods, which are fully data-driven, we choose the physically meaningful quantities \emph{a priori}, and define them as parameters and variables in the proposed reduced-order model.
Subsequently, this qualitative model is made quantitatively predictive by augmenting it with data.
Data assimilation based on the ensemble Kalman filter is performed to achieve a statistically optimal combination of theoretical, computational and experimental knowledge~\cite[e.g.,][]{Evensen2009,Magri2020springer}.
The ensemble Kalman filter is suitable for the treatment of nonlinear dynamics because it is a stochastic technique.
By using a Bayesian framework, the objective of this work is to propose an adaptive reduced-order model that predicts the nonlinear flame dynamics with its uncertainties.
We collect experimental data, and develop a reduced-order model of the base flow and the flame response (Sections~\ref{sec:exp}, \ref{sec:G}).
Using the ensemble Kalman filter, we assimilate the experimental data into the reduced-order model for optimal calibration, and quantify the uncertainties in the flame reponse and the model parameters (Section~\ref{sec:da}).
In Section~\ref{sec:sum}, we summarize the conclusions of this study.


\FloatBarrier


\section{Experiment}
\label{sec:exp}

%

\begin{figure}[ht]
\centering
\includegraphics[width=0.75\columnwidth]{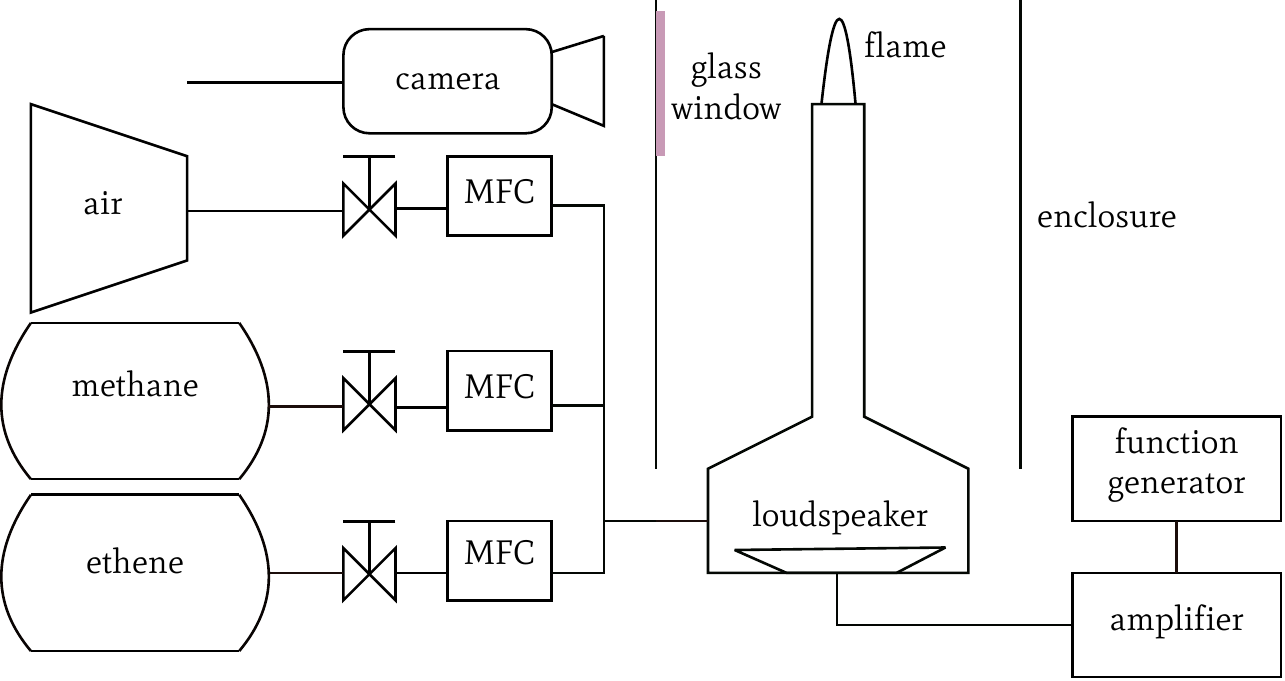}
\caption{Experimental set-up.}
\label{fig:exp:set_up}
\end{figure}

\begin{figure*}[t]
\centering
\includegraphics[height=0.12\linewidth]{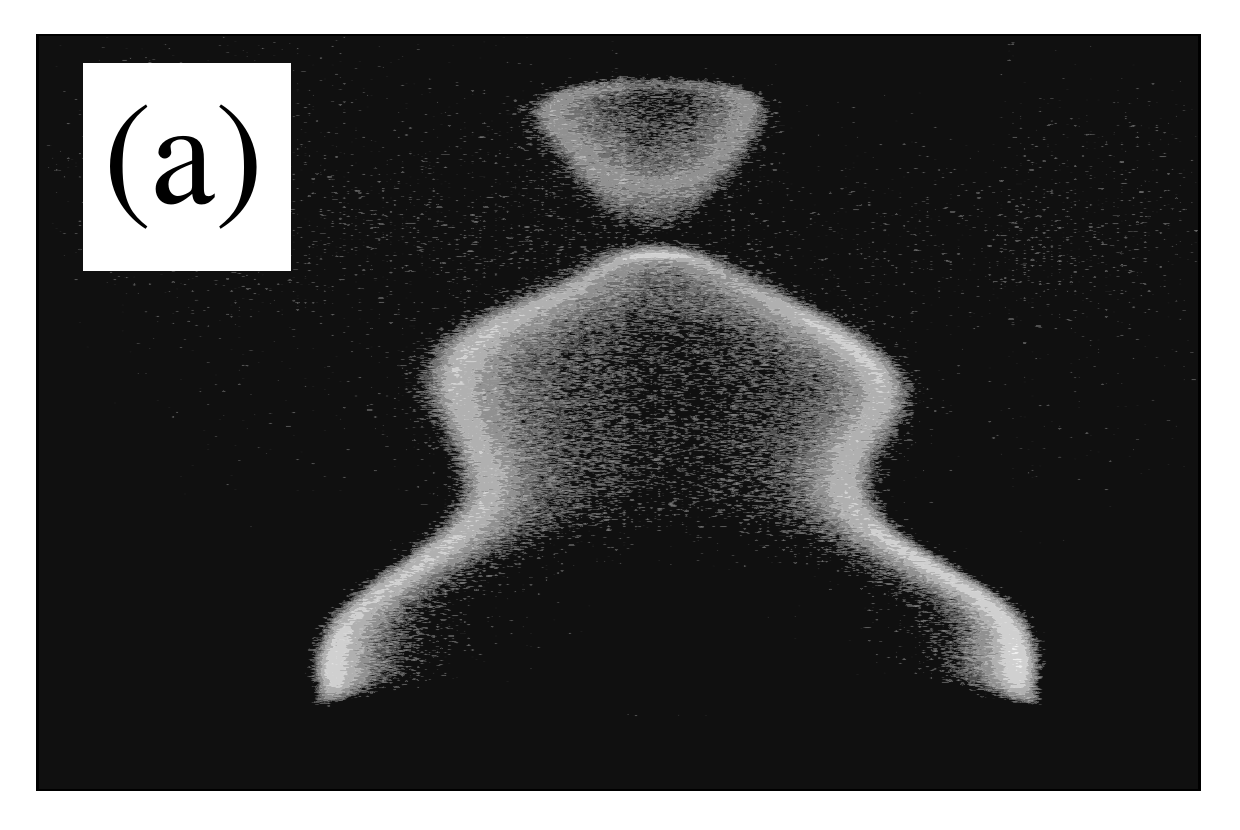}
~
\includegraphics[height=0.12\linewidth]{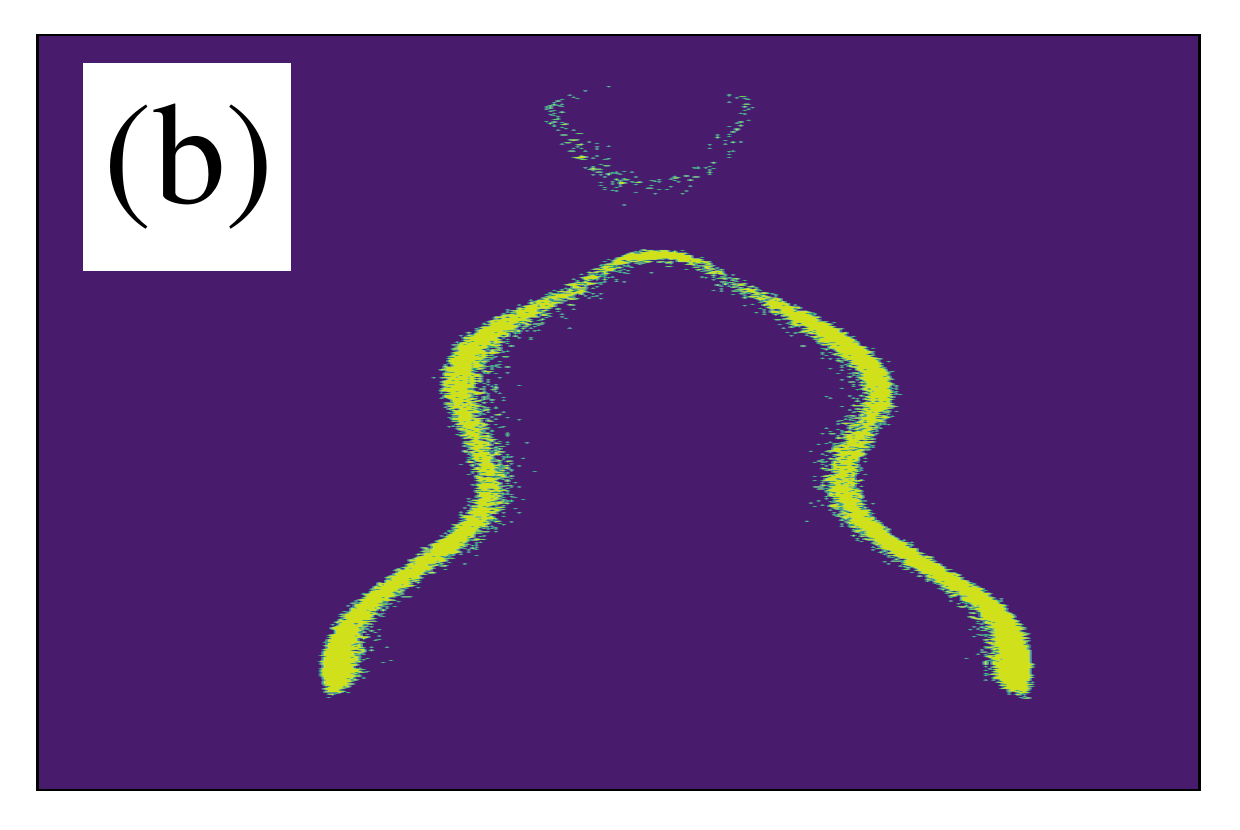}
~
\includegraphics[height=0.12\linewidth]{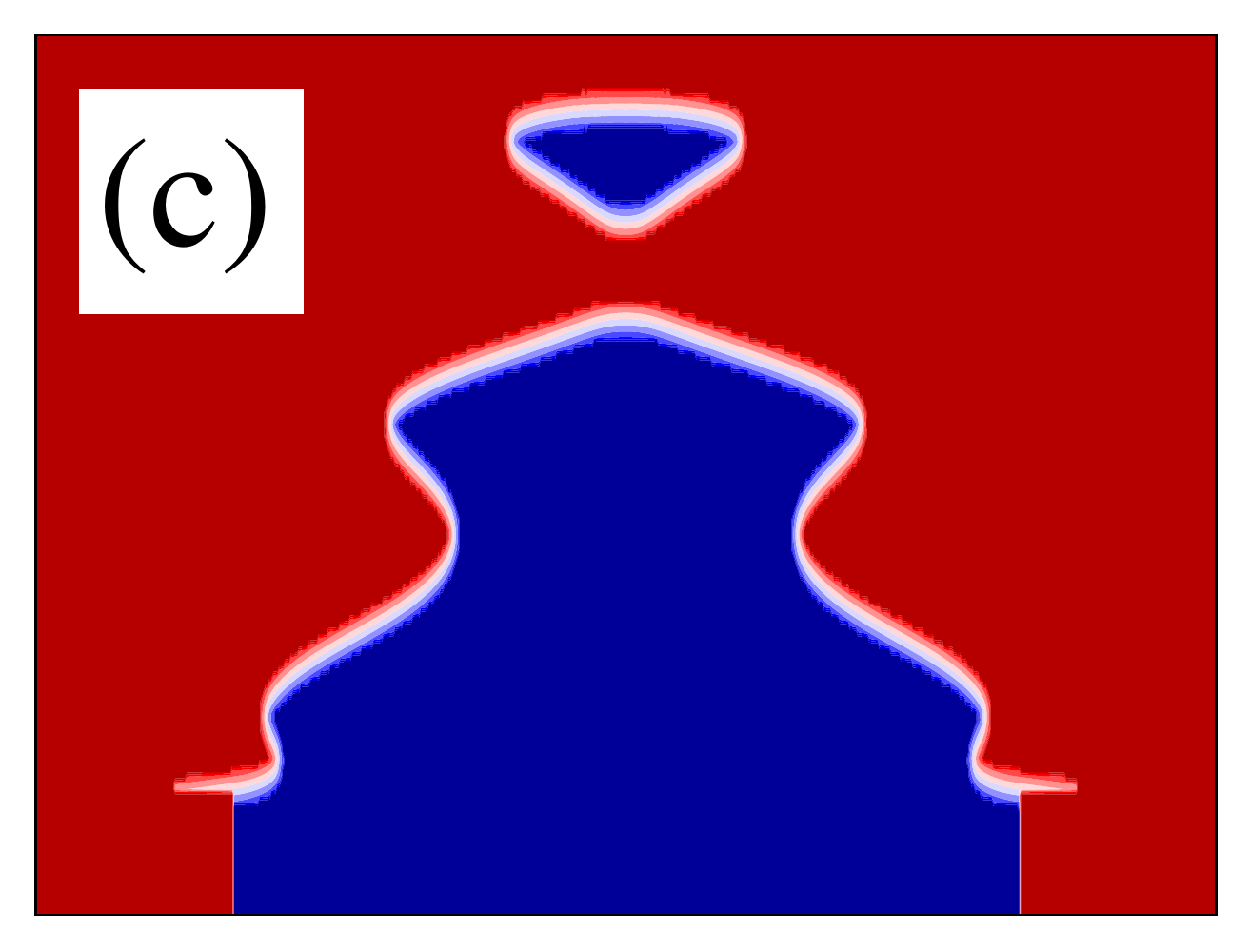}
~
\includegraphics[height=0.12\linewidth]{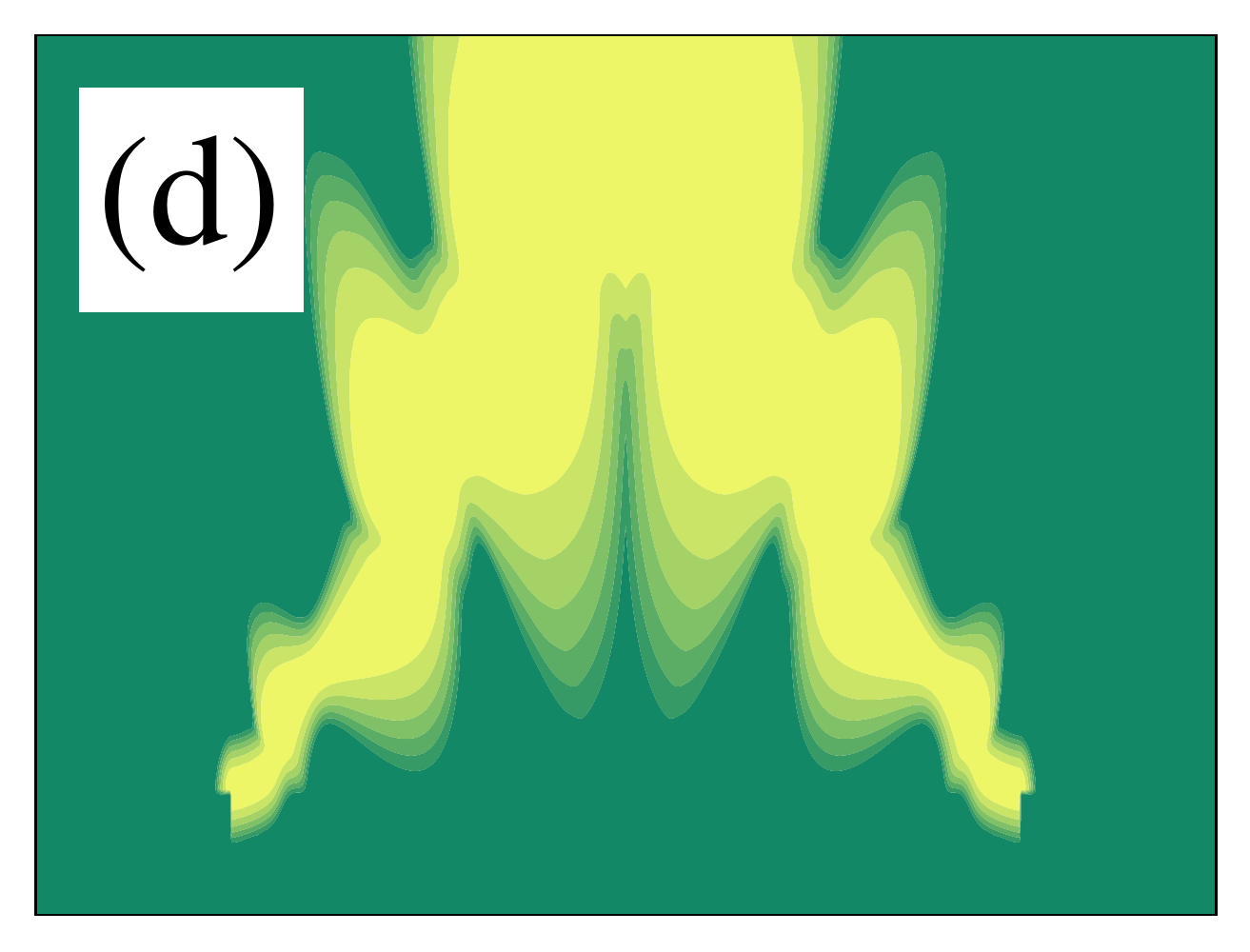}
~
\includegraphics[height=0.12\linewidth]{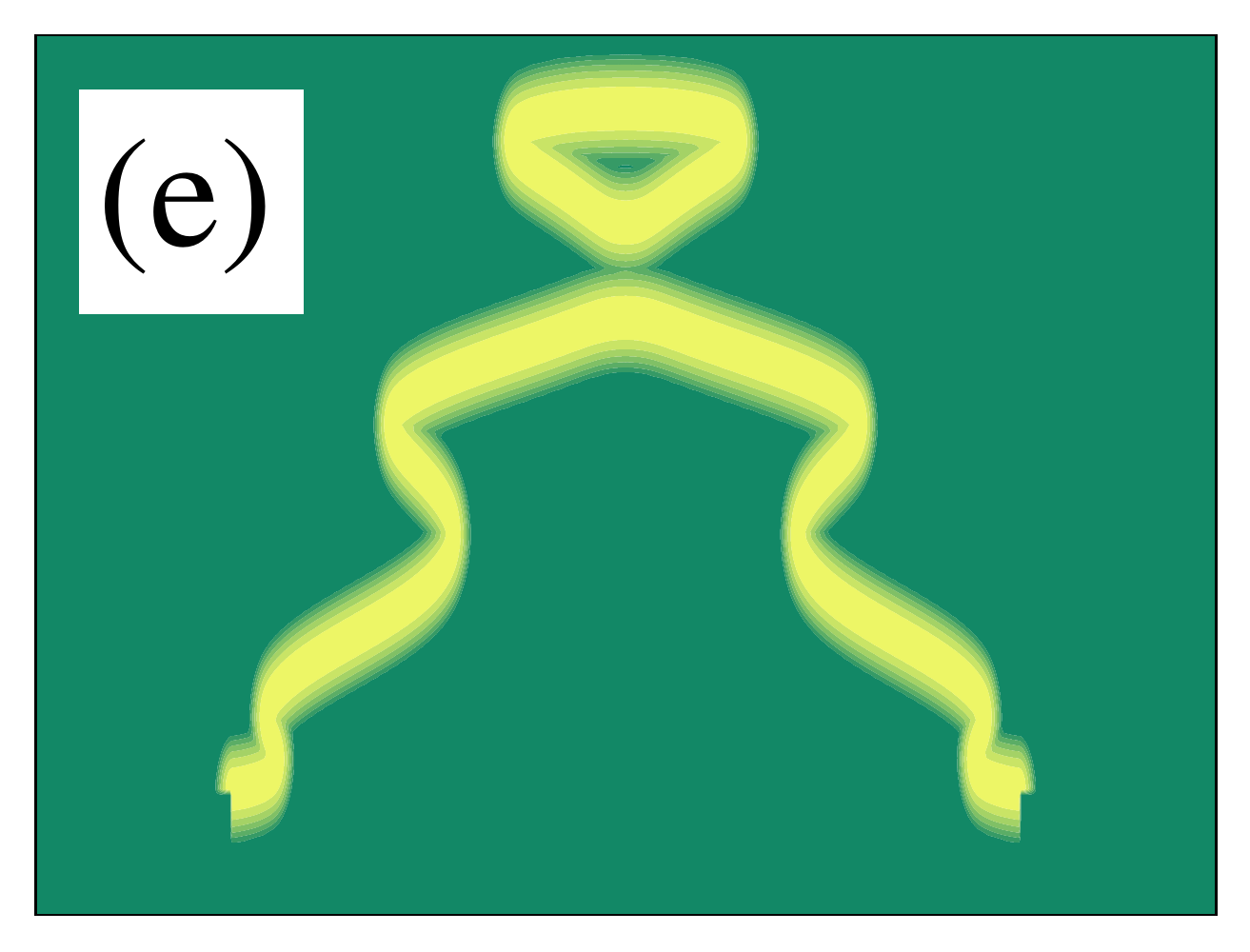}
~
\raisebox{0.005\linewidth}{\includegraphics[height=0.12\linewidth]{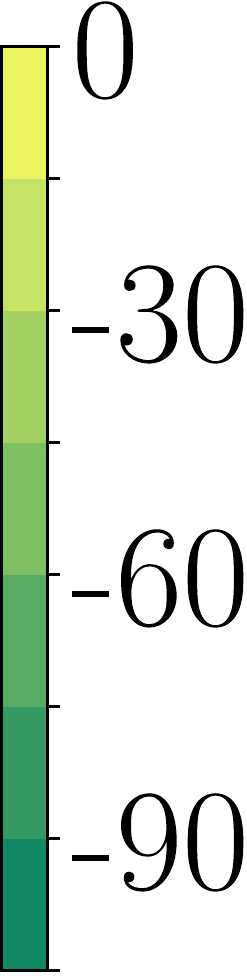}}
\caption{
Experimental, computational and statistical visualisations of premixed flame at reduced aspect ratios.
(a):
Experimental image of acoustic forcing at $200\,\mathrm{Hz}$.
Perturbations of the flame surface form at the base and travel to the tip.
If the amplitude of the perturbation is large enough, a fuel-air pocket pinches off.
(b):
Experimental image after postprocessing.
Pixels associated with the flame surface are colored yellow.
(c):
Snapshot of $G$-equation simulation for $K = 0.5$ and $\varepsilon = 0.36$.
The zero-level set (white) separates the burnt~(red) from the unburnt~(blue) region.
(d):
Snapshot of log-normalized likelihood (Eq.~\eqref{eq:da:sp:loglik}) for initial guess of $K = 0.5$ and $\varepsilon = 0.2$ with 10\,\% standard deviation in each.
Highly likely positions of the flame surface are shown in yellow; less likely positions in green.
(e):
Snapshot of log-normalized likelihood after combined state and parameter estimation.
The model parameters~$K$ and~$\varepsilon$ are optimally calibrated such that they reproduce the propagation of perturbations along the flame surface as well as the formation of pinched-off fuel-air pockets.
The spread of the high-likelihood locations~(yellow) resembles the precision of the edge detection~(b).
The computational and statistical results of panels (c,d,e) are thoroughly explained in Sections~\ref{sec:G}~and~\ref{sec:da}.
}
\label{fig:exp:inout}
\end{figure*}

A schematic view of the experimental set-up is given in Fig.~\ref{fig:exp:set_up}.
The core of the experiment consists of a Bunsen burner inside a tube.
The Bunsen burner consists of a straight metal pipe with an inner diameter of $10\,\mathrm{mm}$.
A tube with an integrated glass window for optical access acts as a cylindrical enclosure to shield the flame from gusts.
Experiments are carried out with a premixed methane-ethene-air mixture with laminar flow rates set to $0.9$, $0.25$ and $8$ normal liters per minute.
The composition of the mixture is controlled with \textsc{Bronkhorst} EL-FLOW mass flow controllers~(MFC).
A loudspeaker, driven by an amplified sinusoidal signal, is mounted upstream of the Bunsen burner for acoustic forcing.
The flame dynamics are recorded with a \textsc{Phantom} V4.2 CMOS camera with a glass lens at a resolution of $1280\times800\,\mathrm{pixels}$ and a frame rate of $2800\,\mathrm{frames}/\mathrm{s}$.
The resolution is sufficiently fine to resolve the flame surface~(Fig.~\ref{fig:exp:inout}a).


In order to perform data assimilation~(Section~\ref{sec:da}), we extract the coordinates of the pixels associated with the flame surface, which is identified by its natural luminescence in the visible range in the experimental images~(Fig.~\ref{fig:exp:inout}a).
A number of edge and ridge detection algorithms are readily available from \textsc{scikit-image}~\cite{Walt2014}, all of which give comparable results.
In Fig.~\ref{fig:exp:inout}b, the result of applying the Sobel filter, which is used in this paper, is shown.




\section{Reduced-order model}
\label{sec:G}

\begin{figure}[ht]
\centering
\includegraphics[width=0.45\columnwidth]{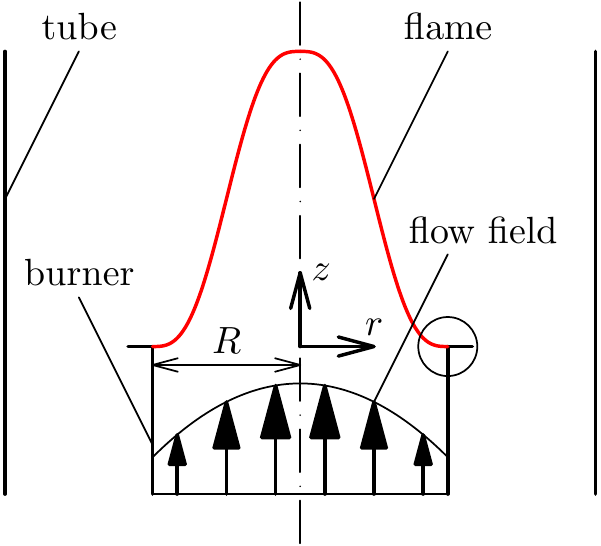}
~
\includegraphics[width=0.45\columnwidth]{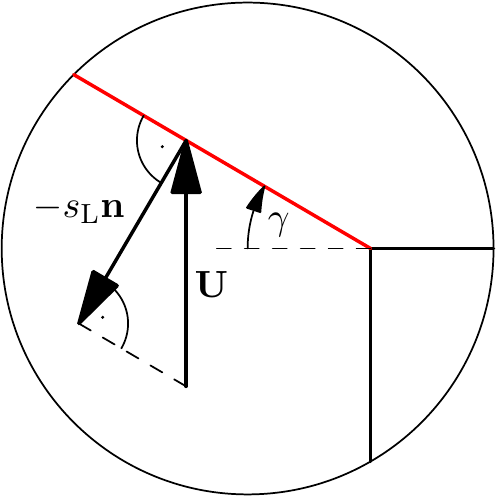}
\caption{
Reduced-order model of the ducted premixed flame (left).
In the absence of velocity perturbations, there is a kinematic balance between the base flow and the flame surface, e.g.\ at the burner lip (right).
}
\label{fig:G:rom}
\end{figure}

The evolution of a premixed flame is modeled here by the kinematics of a surface.
The position of the flame surface is captured by the zero-level set of a continuous scalar variable~$G$.
It is governed by the $G$-equation~\cite{Peters2000}:
\begin{equation}
\pdiff{G}{t} + (\vect{u} - s_\mathrm{L}\vect{n}) \cdot \nabla G = 0 \quad ,
\label{eq:G:G}
\end{equation}
where~$\vect{u}$ is the underlying flow field, $s_\mathrm{L}$ is the laminar flame speed, and~$\vect{n}$ is the unit normal vector.
The underlying flow field~$\vect{u}$ is the superposition of a base flow~$\vect{U}$ (Section~\ref{sec:G:bf}) and a velocity perturbation~$\vect{u'}$ (Section~\ref{sec:G:ac}).
The laminar flame speed is
\begin{equation}
s_\mathrm{L} = s_\mathrm{L}^0 (1 - \kappa \mathcal{L}) \quad ,
\label{eq:G:sL}
\end{equation}
where $s_\mathrm{L}^0$ denotes the adiabatic flame speed.
The Markstein length~$\mathcal{L}$ makes the flame speed a function of the local curvature~$\kappa$.
Normal vector~$\vect{n}$ and curvature~$\kappa$ are given in terms of~$G$:
\begin{equation}
\vect{n} = \frac{\nabla G}{\|\nabla G\|} \quad , \quad \kappa = -\nabla \cdot \vect{n} \quad .
\label{eq:G:geo}
\end{equation}
Fig.~\ref{fig:G:rom} shows a schematic of this reduced-order model.

The $G$-equation is numerically solved using a narrow-band level-set method with distance reinitialization.
The computational domain is discretized using a fifth-order, weighted essentially non-oscillatory~(WENO) scheme in space and a third-order, total variation diminishing~(TVD) Runge-Kutta scheme in time.
At the burner lip, a rotating boundary condition is used~\cite{Waugh2013}.
For details on the $G$-equation solver, the reader is referred to~\cite{Hemchandra2009}.


\subsection{Base flow}
\label{sec:G:bf}

For a time-independent base flow~$\vect{u} = \vect{U}$, the $G$-equation (Eq.~\ref{eq:G:G}) becomes
\begin{equation}
\vect{U} \cdot \vect{n} - s_\mathrm{L}^0 (1 - \kappa \mathcal{L}) = 0 \quad .
\label{eq:G:bf:G}
\end{equation}
For an axisymmetric burner, we parametrize the zero-level set, i.e.\ the flame surface, by setting its height~$h$ above the burner outlet as a function of the radius~$r$:
\begin{equation}
G(r, z=h(r)) = 0 \quad , \quad 0 \leq r \leq R \quad .
\end{equation}
The normal vector~$\vect{n}$ and the curvature~$\kappa$ (Eq.~\ref{eq:G:geo}) in terms of~$h$ are
\begin{equation}
\vect{n} = \frac{1}{\left[1 + h'(r)^2\right]^{0.5}}
\begin{pmatrix}
-h'(r) \\
1
\end{pmatrix} \quad ,
\label{eq:G:bf:geo1}
\end{equation}
\begin{equation}
\kappa = \frac{h''(r)}{\left[1 + h'(r)^2\right]^{1.5}} + \frac{h'(r)}{r\left[1 + h'(r)^2\right]^{0.5}} \quad .
\label{eq:G:bf:geo2}
\end{equation}
Substituting Eqs.~\ref{eq:G:bf:geo1} and~\ref{eq:G:bf:geo2} into Eq.~\ref{eq:G:bf:G} gives a one-dimensional boundary value problem with $h'(0) = 0$ and $h(R) = 0$.

For simplicity, we assume that the base flow~$\vect{U}$ is only a function of the radius~$r$, but not of the height~$h$.
While the conditions at the burner outlet are theoretically known to be those of a Poiseuille-type pipe flow for a given mass flow rate, we introduce two additional parameters to account for the imperfections of this reduced-order model.
Firstly, the velocity profile deviates from that of a Poiseuille flow towards a uniform flow the further we move away from the burner outlet~\cite{Schlichting2006}.
Secondly, we observe that, as the base-flow speed vanishes near the burner wall due to the no-slip condition, the flame speed also decreases due to heat loss to the burner wall~\cite{Cuquel2013}, which is not properly modeled by the constant adiabatic flame speed~$s_\mathrm{L}^0$ (Eq.~\eqref{eq:G:sL}).
Hence, we introduce two additional parameters to the reduced-order model of the base flow:
The aspect ratio~$\beta$ gives the ratio between the flame length and the burner radius in a uniform flow without curvature effects:
\begin{equation}
\left(\frac{\bar{U}}{s_\mathrm{L}^0}\right)^2 = \beta^2 + 1 \quad ,
\label{eq:G:bf:beta}
\end{equation}
where~$\bar{U}$ denotes the mean flow speed.
The shape parameter~$\alpha$ linearly determines the velocity profile such that the mass flow rate is conserved:
\begin{equation}
\frac{U}{\bar{U}} = 1 + \alpha \left[1 - 2 \left(\frac{r}{R}\right)^2\right] \quad ,
\label{eq:G:bf:alpha}
\end{equation}
where $\alpha = 0$ corresponds to a uniform flow, and $\alpha = 1$ corresponds to a Poiseuille flow.
In summary, the base-flow model has three parameters~$\boldsymbol{\theta}_\mathrm{bf}$ (`base flow'): the shape parameter~$\alpha$, the aspect ratio~$\beta$ and the Markstein length~$\mathcal{L}$.
Note that $\beta$ replaces the parameter $s_\mathrm{L}^0/\bar{U}$ due to Eq.~\eqref{eq:G:bf:beta}.

For a given set of parameters, we solve the boundary value problem (Eqs.~\eqref{eq:G:bf:G}-\eqref{eq:G:bf:geo2}) by using a finite-difference method and a Newton-Raphson solver~\cite{Ascher1995}.
In iteration step~$k$, the residual $\vect{R}^k = \vect{R}(\vect{h}^k)$ is computed by evaluating the left-hand side of Eq.~\eqref{eq:G:bf:G} at every grid point.
The discretization of the height, $h$, is represented by $\mathbf{h}$.
The Jacobian $\vect{J}^k = \vect{J}(\vect{h}^k)$ is computed by applying the chain rule to differentiate $\vect{R}^k$ with respect to~$\vect{h}^k$.
Updates are performed by repeatedly solving
\begin{equation}
\vect{J}^k \Delta{\vect{h}^k} = \vect{R}^k \quad \implies \quad \vect{h}^{k+1} = \vect{h}^k - \Delta{\vect{h}^k} \quad .
\label{eq:G:bf:newton}
\end{equation}

To infer the values of the parameters, we embed the boundary value problem into a least-squares problem with a cost functional~$\mathfrak{E}$:
\begin{equation}
\mathfrak{E} = \sum_{m=1}^M{\left[z_m - L_m(\vect{h}(\boldsymbol{\theta}_\mathrm{bf}))\right]^2} \quad ,
\end{equation}
where $L_m$ is a suitable linear interpolation operator for the $m$-th measurement $(r_m, z_m)$ of the flame surface.
This optimization problem is solved by line search~\cite{Nocedal2006}.
The sensitivity of the cost functional~$\mathfrak{E}$ to the parameters~$\boldsymbol{\theta}_\mathrm{bf}$ is calculated using the adjoint variable~$\boldsymbol{\lambda}$~\cite{Gunzburger2002}:
\begin{equation}
\boldsymbol{\lambda}^\mathrm{T} \pdiff{\vect{R}}{\vect{h}} = \pdiff{\mathfrak{E}}{\vect{h}} \quad \implies \quad \tdiff{\mathfrak{E}}{\boldsymbol{\theta}_\mathrm{bf}} = \boldsymbol{\lambda}^\mathrm{T} \pdiff{\vect{R}}{\boldsymbol{\theta}_\mathrm{bf}} \quad .
\end{equation}

\begin{figure}[ht]
\centering
\includegraphics[width=0.75\columnwidth]{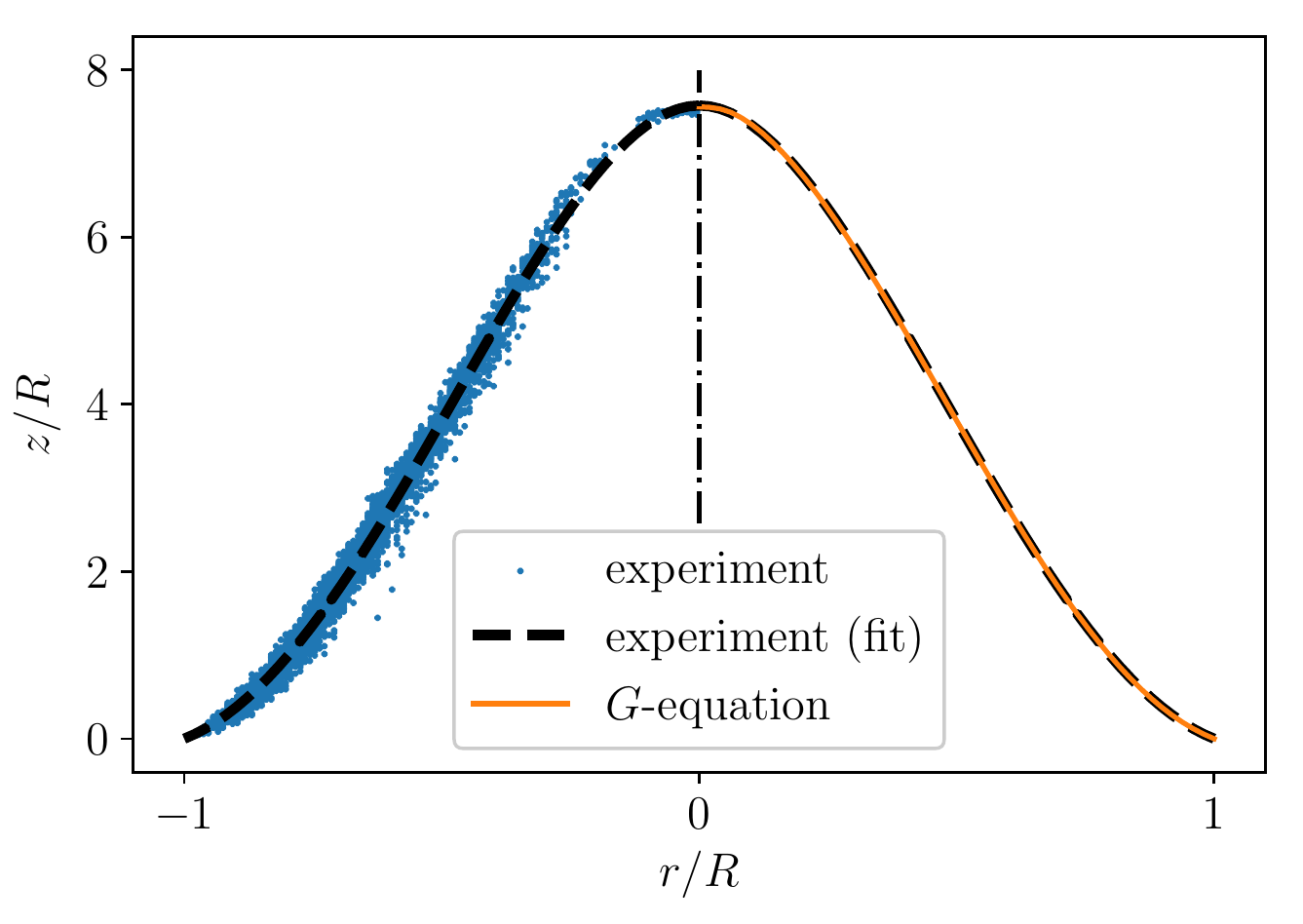}
\caption{
Edge detection (blue dots, left) and least-squares fit from base-flow model (orange line, right).
For comparison, the fourth-order polynomial fit $z/R = a_4(r/R)^4 + a_3(r/R)^3 + a_2(r/R)^2 - (a_4 + a_3 + a_2)$ respecting the boundary conditions is shown (black dashed line, both left and right).
The base-flow model reproduces the flame surface observed in the experimental images.
}
\label{fig:G:bf:line_search}
\end{figure}

In Fig.~\ref{fig:G:bf:line_search}, the results are shown for $\alpha = 0.84$, $\beta = 15.1$ and $\mathcal{L} = 3\,\mathrm{mm}$.
The base-flow model agrees with the experiment.
Furthermore, the base-flow model is confirmed using \textsc{Cantera} simulations~\cite{Goodwin2018}, which provide $\beta \approx 15.8$, based on the calculated adiabatic flame speed as well as the mass flow rate set in the experiment, and a flame thickness of $1.2\,\mathrm{mm}$, the latter on the same order of magnitude as the inferred Markstein length, in agreement with the theory~\cite{Clavin1982, Matalon1982}.
Finally, $\alpha = 0.84$ indicates a velocity profile close to Poiseuille flow as expected.
\revis{
Therefore, our model with three parameters covers a variety of base flows on physical grounds. Using a more complex base-flow model requires no conceptual changes to the data assimilation framework.
}

\subsection{Flame response}
\label{sec:G:ac}

For the forcing of the premixed flame, the radial and axial components of the velocity perturbation~$\mathbf{u'}$ are~\cite{Kashinath2013}
\begin{equation}
\frac{u_r'}{\bar{U}} = -\frac{\varepsilon\pi f K r}{\bar{U}} \cos\left(2\pi f \left(\frac{Kz}{\bar{U}} - t\right) \right) \quad ,
\label{eq:G:ac:u_r}
\end{equation}
\begin{equation}
\frac{u_z'}{\bar{U}} = \varepsilon \sin\left(2\pi f \left(\frac{Kz}{\bar{U}} - t\right) \right) \quad ,
\label{eq:G:ac:u_z}
\end{equation}
where $u_r'$ and $u_z'$ satisfy the continuity equation.
The forcing has a frequency~$f$ and an amplitude~$\varepsilon$.
The non-dimensional parameter~$K$ is the ratio between the mean flow speed and the phase speed at which perturbations on the flame surface travel in the axial direction.
The model parameters~$\boldsymbol{\theta}_\mathrm{fr}$ (`flame response') are $K$ and~$\varepsilon$.
In Fig.~\ref{fig:exp:inout}c, the result from a simulation based on our reduced-order model with parameters chosen to qualitatively match Fig.~\ref{fig:exp:inout}a,b is shown.



In preparation for data assimilation (Section~\ref{sec:da}), the $G$-equation has to be synchronized with the experimental images.
This is not straightforward because we control the phase of the velocity perturbations in the $G$-equation (Eqs.~\eqref{eq:G:ac:u_r}, \eqref{eq:G:ac:u_z}), whereas the experimental images only depict the flame surface without any velocity information.
Therefore, we derive an analytical relationship between the velocity perturbations and the motion of the flame surface at the base of the flame.
As shown in Fig.~\ref{fig:G:rom} (right), the relationship between the base flow (Eq.~\eqref{eq:G:bf:alpha}) and the flame surface (Eq.~\eqref{eq:G:bf:beta}) at the burner lip ($r=R$, $z=0$), neglecting curvature effects, is
\begin{equation}
\cos(\gamma)
= \left.\frac{s_\mathrm{L}}{U}\right|_R
\approx \frac{s_\mathrm{L}^0}{U(R)}
\approx \frac{1}{\beta(1 - \alpha)} \quad ,
\end{equation}
where the last approximation is justified for $\beta = 15.1 \gg 1$ as observed in the experiment.
The normal vector~$\mathbf{n}$ is
\begin{equation}
n_r
\approx 1 \quad , \quad
n_z = \frac{1}{\beta (1 - \alpha)} \quad .
\end{equation}
We consider a small-amplitude perturbation~$\gamma'$ around the angle $\gamma$ as a result of the corresponding velocity perturbation~$\mathbf{u'}$:
\begin{align}
\mathbf{u'} \cdot \mathbf{n}
\approx -\varepsilon\pi fKR \sin\left(2\pi f t + \Delta\varphi\right) \quad ,
\label{eq:G:ac:gamma_dot}
\end{align}
where it is assumed that $\left(\pi fKR \beta(1 - \alpha) / \bar{U}\right)^2 \gg 1$ and $\tan(\Delta\varphi) = \pi fKR \beta(1 - \alpha) / \bar{U}$, which is justified by the inferred values for the model parameters~$\boldsymbol{\theta}_\mathrm{bf}$ and the judicious choice of frequencies~$f$ in Section~\ref{sec:da}.
By observing the motion of the flame surface near the burner lip, $\Delta\varphi$ is calibrated to synchronize the $G$-equation with the experimental images.

Before turning to the data-driven estimation of~$K$ and~$\varepsilon$ in the next section, we summarize the \emph{a-priori} insights about the model parameters.
Under the assumption that the velocity perturbation felt at the base of the flame, i.e.\ $\dot{\gamma}'$, only depends on the volume of the loudspeaker, it follows from Eq.~\eqref{eq:G:ac:gamma_dot} that the amplitude, $\varepsilon$, is inversely proportional to the frequency, $f$.
Consequently, Eq.~\eqref{eq:G:ac:gamma_dot} implies a low-pass filter for the flame response~\cite{Ducruix2000}.
The assumption that~$K$ only depends on the base flow, not the forcing frequency, is expected to be valid at small amplitudes, as shown by linear stability analysis~\cite{Landau1988}.
The behavior at larger amplitudes can be investigated with the approach in this paper.
While~$K \approx 1$ is reasonable for open flames in quiescent environments~\cite{Schuller2002, Schuller2003}, we additionally have to take into account the entrainment due to the buoyancy-driven flow surrounding the burner as well as the confinement due to the enclosing tube.
Hence we anticipate a frequency-independent phase speed for the velocity perturbation with $K < 1$.



\section{Data assimilation}
\label{sec:da}

The Kalman filter provides a statistically optimal estimate~$\psi^a$ (`analysis') of the unknown state~$\psi$ of a system from a model prediction~$\psi^f$ (`forecast') and experimental observations~$\mathbf{d}$~\cite{Evensen2009}.
The model prediction is mapped from its state space to the observation space through a measurement operator~$\mathbf{M}$.
The prediction uncertainties and the experimental errors are represented by covariance matrices~$\mathbf{C}_{\psi\psi}^f$ and~$\mathbf{C}_{\epsilon\epsilon}$, respectively.

The application of the Kalman filter to the proposed reduced-order model is challenging for at least two reasons:
Firstly, the $G$-equation is highly nonlinear, which is exemplified by the occurrence of cusps and pinched-off fuel-air pockets.
This complicates the treatment of the time-dependent covariance matrix~$\mathbf{C}_{\psi\psi}^f$.
Secondly, the proposed reduced-order model has $\mathcal{O}(10^5)$~degrees of freedom after discretization, which makes the computation and inversion of covariance matrices computationally intractable.
To make the analysis statistically and computationally feasible, we instead generate an ensemble of $N$~model predictions~$\psi_i^f$ with $i=1,\dots,N$.
This variation of the Kalman filter, the ensemble Kalman filter, gives for~$\psi_i^a$ and its statistics~\cite{Evensen2009, Yu2019}:
\begin{equation}
\psi_i^a = \psi_i^f + \left(\mathbf{M}\mathbf{C}_{\psi\psi}^f\right)^\mathrm{T}\left[\mathbf{C}_{\epsilon\epsilon}+\mathbf{M}\mathbf{C}_{\psi\psi}^f\mathbf{M}^\mathrm{T}\right]^{-1}\left(\mathbf{d}-\mathbf{M}\psi_i^f\right) \quad ,
\label{eq:da:analysis}
\end{equation}
\begin{equation}
\bar{\psi} = \frac{1}{N}\sum_{i=1}^N{\psi_i} \quad , \quad \mathbf{C}_{\psi\psi} = \frac{1}{N-1}\sum_{i=1}^N{\left(\psi_i - \bar{\psi}\right)\left(\psi_i - \bar{\psi}\right)^\mathrm{T}} \quad .
\label{eq:da:statistics}
\end{equation}
Finally, the question remains as to whether the (ensemble) Kalman filter preserves the kinematic nature of our reduced-order model.
The choice of a suitable state space, i.e.\ the construction of~$\psi$ from~$G$, is crucial and not straightforward.
Hence, we adopt the level-set data assimilation framework developed by Yu et al.~\cite{Yu2019}, which is based on the Hamilton-Jacobi formalism.



In state estimation, the ensemble~$\psi_i$ represents one realization from a probability distribution in~$\psi$ with mean~$\bar{\psi}$ and covariance matrix~$\mathbf{C}_{\psi\psi}$.
As such, the marginal probability distribution in the $k$-th entry of~$\psi$ is given by the mean~$\bar{\psi}[k]$ and the variance~$\mathbf{C}_{\psi\psi}[k, k]$.
Consequently, the likelihood of finding the flame surface at the location corresponding to the $k$-th entry of~$\psi$, regardless of the position of the flame surface elsewhere, is~\cite{Yu2019}
\begin{equation}
p[k] = \frac{1}{\sqrt{2\pi\mathbf{C}_{\psi\psi}[k, k]}}\exp\left(-\frac{\bar{\psi}[k]^2}{2\mathbf{C}_{\psi\psi}[k, k]}\right) \quad .
\label{eq:da:sp:lik}
\end{equation}
Alternatively, the log-normalized likelihood is given by
\begin{equation}
\log\left(\frac{p[k]}{p_0[k]}\right) = -\frac{\bar{\psi}[k]^2}{2\mathbf{C}_{\psi\psi}[k, k]} \quad ,
\label{eq:da:sp:loglik}
\end{equation}
where $\log(p/p_0) = 0$ identifies the most likely position of the flame surface.
In Fig.~\ref{fig:exp:inout}d, the log-normalized likelihood is visualized for an educated guess of $K\approx0.5$ and $\varepsilon\approx0.2$ before any data assimilation.
The position of the flame surface becomes highly uncertain for just a modest amount of standard deviation in the model parameters.

\begin{figure}[ht]
\centering
\includegraphics[width=0.75\columnwidth]{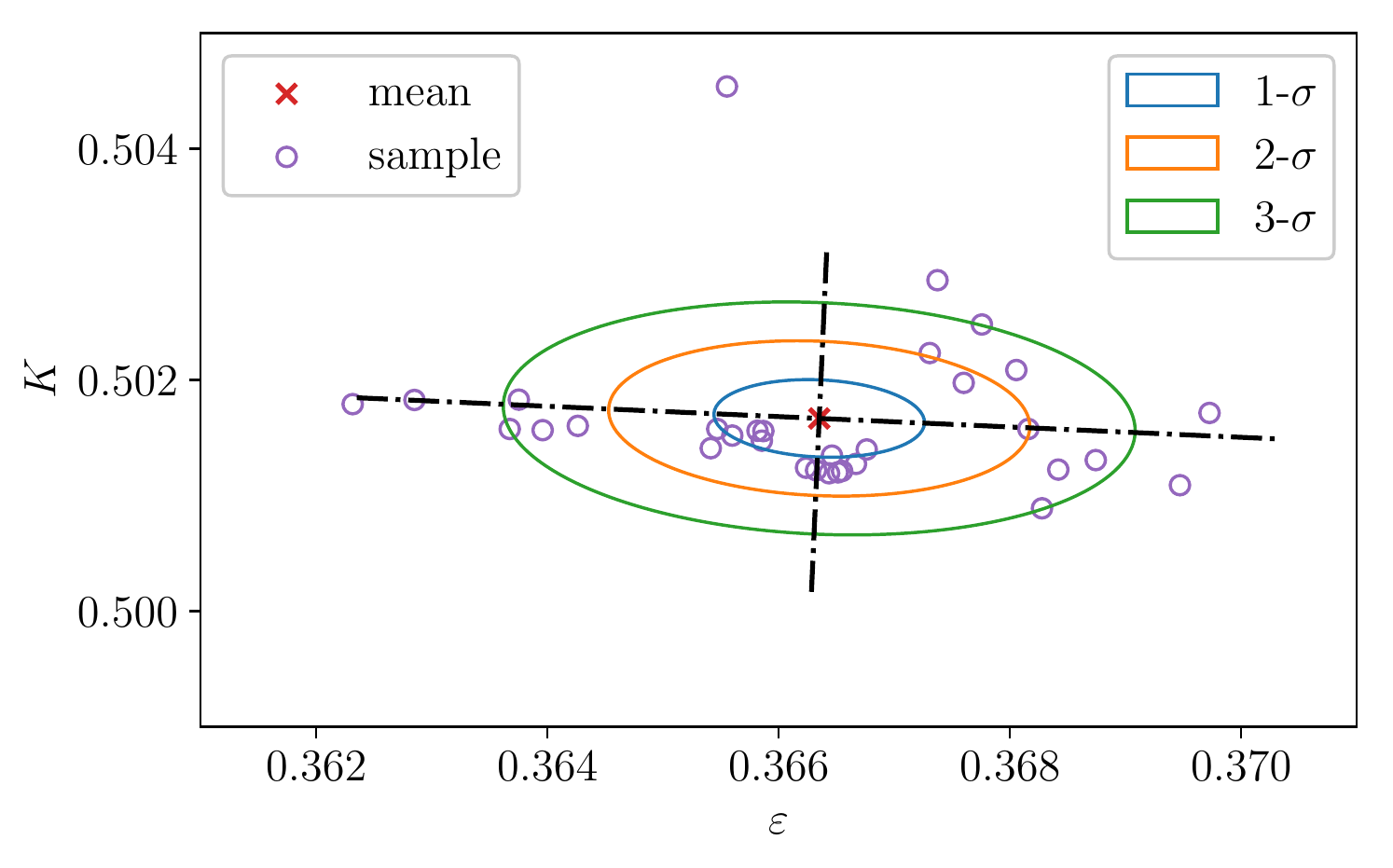}
\caption{
Sample of~$K$ and~$\varepsilon$ and reconstructed probability distribution after combined state and parameter estimation for $f = 200\,\mathrm{Hz}$.
The 1-, 2- and 3-$\sigma$~(blue/orange/green, respectively) confidence regions correspond to 39, 86 and 99\,\% probabilities of sampling a given set of parameters, respectively.
}
\label{fig:da:sp:a_posteriori:param}
\end{figure}

For combined state and parameter estimation, an augmented $\widetilde{\psi}$ is obtained by appending the model parameters~$\boldsymbol{\theta}_\mathrm{fr}$, i.e.~$K$ and~$\varepsilon$, to~$\psi$, and applying the ensemble Kalman filter to $\widetilde{\psi}$~\cite{Yu2019}.
In Fig.~\ref{fig:exp:inout}e, the results are shown for the same initial guess of~$K$ and~$\varepsilon$ as in Fig.~\ref{fig:exp:inout}d.
In comparison, the identification of high-likelihood positions of the flame surface has significantly improved after combined state and parameter estimation.
In Fig.~\ref{fig:da:sp:a_posteriori:param}, the joint probability distribution in~$K$ and~$\varepsilon$ is visualized.
It is computed by marginalizing $\psi$ from the probability distribution in $\widetilde{\psi}$.
Although the means have the same order of magnitude, the standard deviation in~$K$ is three times smaller.
The parameter~$K$ is easier to infer because the pinch-off timing strongly depends on $K$ and is captured accurately by the proposed data assimilation method.
Moreover, $K$ and~$\varepsilon$ are only weakly correlated, which confirms their distinct roles within the proposed reduced-order model.


To assess the proposed reduced-order model over a range of operating conditions, combined state and parameter estimation is performed for multiple frequencies ($f = 200 - 400\,\mathrm{Hz}$).
Due to the low-pass nature of premixed flames, low frequency forcing leads to large fuel-air pockets, which form cusps and pinch off, while high frequency forcing leads to small perturbations which travel downstream without cusps.
\revis{The range of forcing frequencies is chosen in order to avoid flame blow-off at lower frequencies as well as vanishing flame response at higher frequencies.}
An ensemble of 32~$G$-equation simulations with the same initial condition is considered each time.
The model parameters~$K$ and~$\varepsilon$ are sampled from independent normal distributions based around educated guesses of their values with 10\,\% standard deviation in each.
At first, the $G$-equation simulations are solved without data assimilation to make sure that the dynamics are consistent with the parameters and are free of transient effects.
After 10~periods of forcing, the postprocessed experimental images ($f = 2800\,\mathrm{Hz}$) are assimilated for 5~periods.
The covariance matrix~$\mathbf{C}_{\epsilon\epsilon}$, which represents the experimental errors, is a diagonal matrix with~$\sigma_\epsilon^2$ on its diagonal.
The choice of~$\sigma_\epsilon = 1\,\mathrm{mm}$ is based on the thickness of the detected edges from postprocessing~(Fig.~\ref{fig:exp:inout}b).
\revis{
In general, the precision of the experimental data does not affect the accuracy of the ensemble in the long run, but it reduces the uncertainty in the ensemble overall \cite{Yu2019}.
}
Finally, the $G$-equation simulations are solved for another 5~periods to observe the performance with optimally calibrated model parameters and without any data assimilation.
\revis{
The overall computational time is less than 40 minutes for 20 periods on a cluster node with 1 processor per $G$-equation simulation.
This includes the very frequent output of solution files in order to study time series and statistics for this paper.
This is marginally longer than a single simulation of the $G$-equation as the ensemble Kalman filter is parallel by design.
}

\begin{figure}[ht]
\centering
\includegraphics[width=0.75\columnwidth]{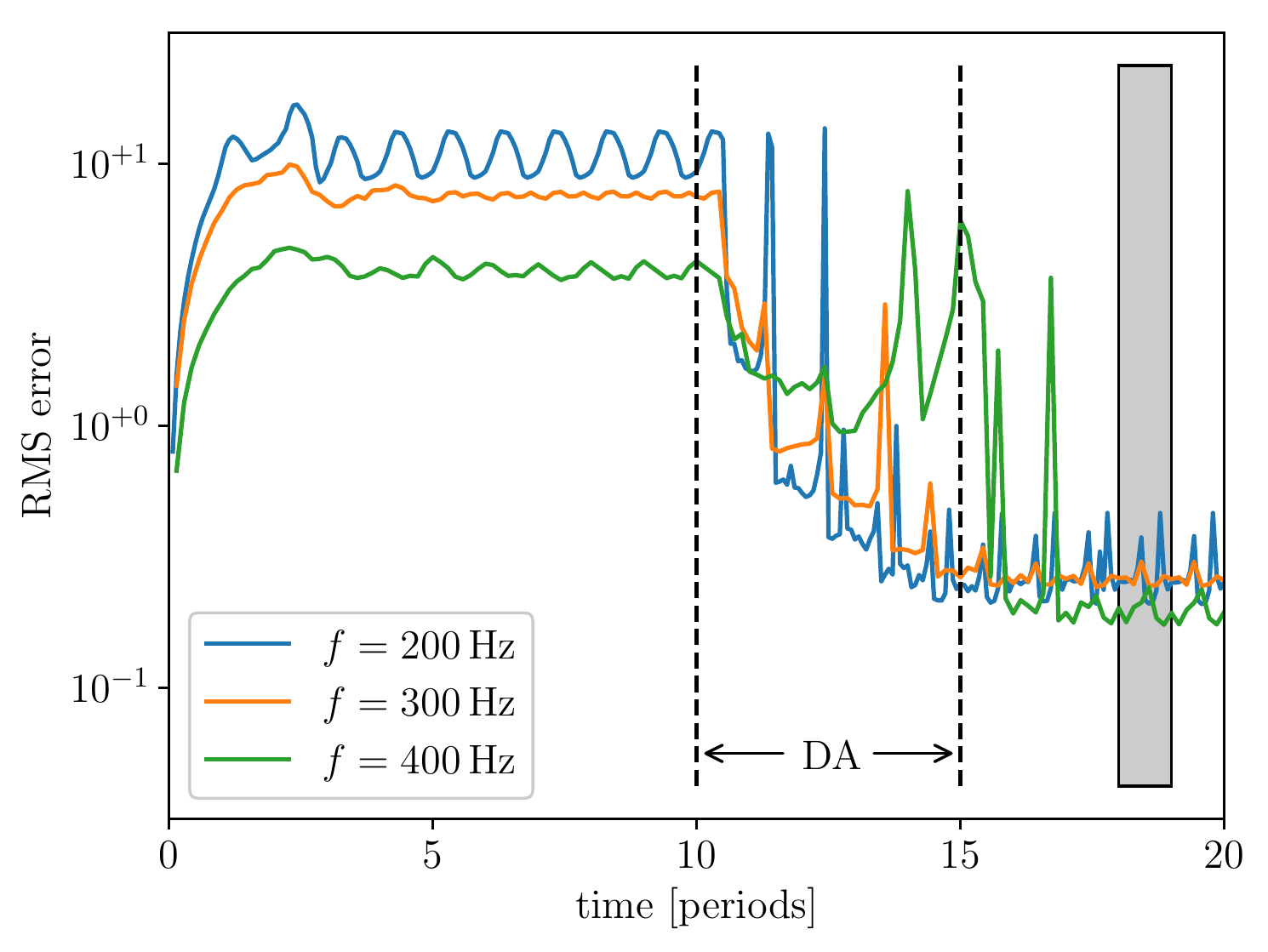}
\caption{
Root-mean-square~(RMS) error for forcing at 200, 300 and 400\,Hz~(blue/orange/green, respectively).
Data assimilation~(DA) takes place between 10 and 15 periods.
The grey window \revis{is representative of} the period depicted in Fig.~\ref{fig:res:heat}.
}
\label{fig:res:s}
\end{figure}

\begin{figure}[ht]
\centering
\includegraphics[width=0.75\columnwidth]{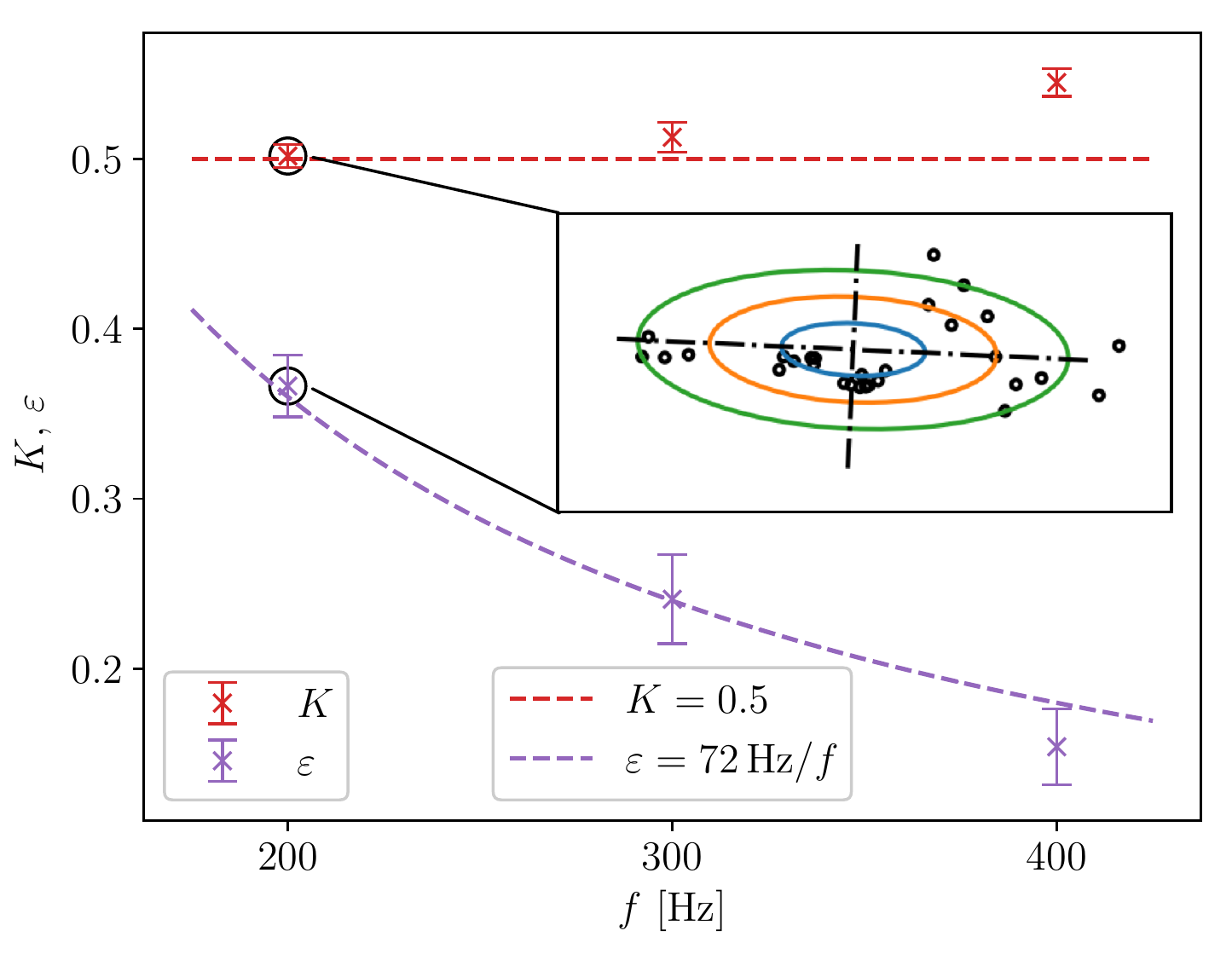}
\caption{
Optimally calibrated estimates and their uncertainties for~$K$~(red) and~$\varepsilon$~(purple).
The 10-$\sigma$ (chosen for clearer visualisation) confidence intervals are computed by marginalizing the corresponding joint probability distributions (Fig.~\ref{fig:da:sp:a_posteriori:param}).
Every joint probability distribution is reconstructed from an ensemble of 32 $G$-equation simulations, 96 simulations in total.
The dashed lines show the behavior estimated from theory (Section~\ref{sec:G:ac}) for $K$ (constant) and $\varepsilon$ (inversely proportional to~$f$).
}
\label{fig:res:p}
\end{figure}

In Fig.~\ref{fig:res:s}, the root-mean-square~(RMS) error,
\begin{equation}
\mathrm{RMS~error} = \sqrt{\frac{1}{N-1}\sum_{i=1}^N{\left(\psi_i - \bar{\psi}\right)^\mathrm{T}\left(\psi_i - \bar{\psi}\right)}} \quad ,
\end{equation}
is plotted over time for combined state and parameter estimation at 200, 300 and 400\,Hz.
Within 5~periods of forcing, the dynamics for the sampled sets of parameters fully form.
Between 10 and 15~periods, the ensemble Kalman filter gradually improves the estimates by up to two orders of magnitude.
After 15~periods, the dynamics adapt to the optimally calibrated model parameters with low uncertainty.
\revis{The remaining spurious oscillations are the result of noise in the experimental data.}
In Fig.~\ref{fig:res:p}, the optimally calibrated model parameters and their confidence intervals are shown for $f=200-400\,\mathrm{Hz}$.
In agreement with theory, the model parameter~$K$ remains nearly constant while the model parameter~$\varepsilon$ decreases in inverse proportion to the frequency~$f$~(Section~\ref{sec:G:ac}).
\revis{
The state of the system strongly varies with the operating conditions because of nonlinear effects. One the other hand, the model and its parametrization either  follow a certain scaling or are constant. (This is because the proposed data-driven reduced-order model is physics-based, as opposed to traditional machine learning algorithms that are physics-blind.) 
This is clearly the case for the model parameters~$K$ and~$\varepsilon$ in Fig.~\ref{fig:res:p}.
Therefore, we expect the reduced-order model to interpolate well despite a limited amount of data.
}
The confidence intervals at the different frequencies are of comparable height.
In agreement with Fig.~\ref{fig:da:sp:a_posteriori:param}, $K$ is significantly more certain than~$\varepsilon$.

\begin{figure*}[t]
\centering
\raisebox{0.0575\linewidth}{200\,Hz}
\quad
\includegraphics[height=0.125\linewidth]{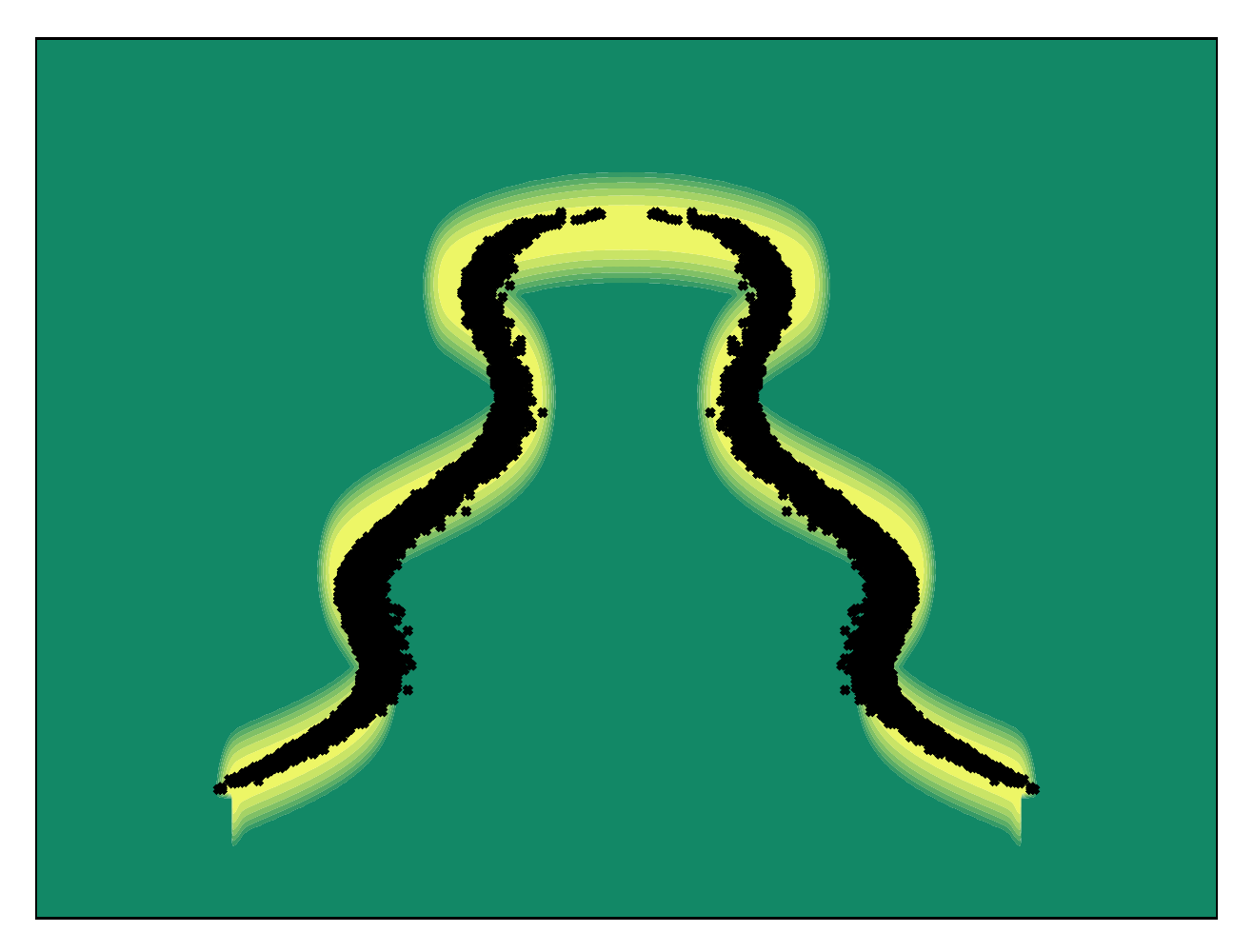}
~
\includegraphics[height=0.125\linewidth]{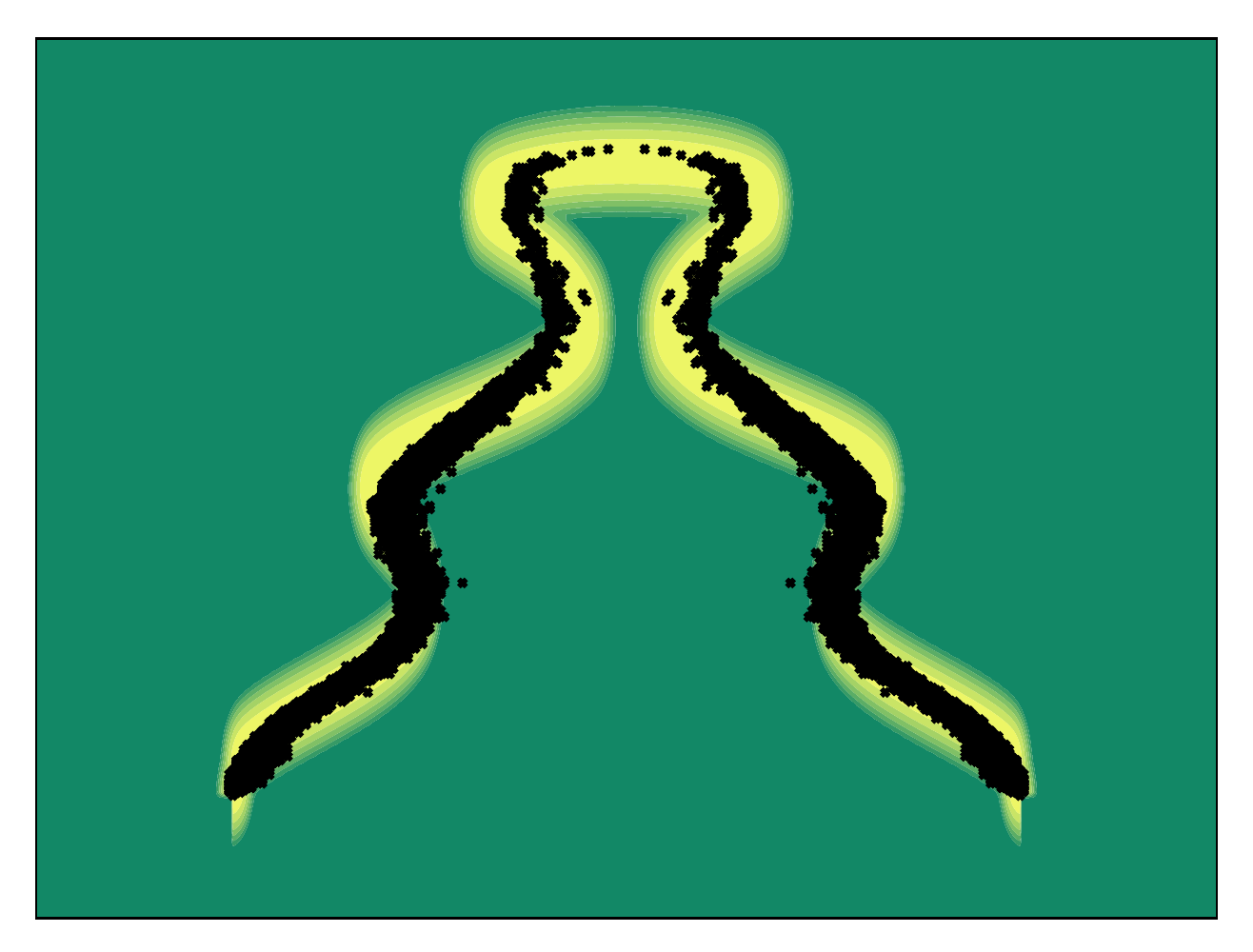}
~
\includegraphics[height=0.125\linewidth]{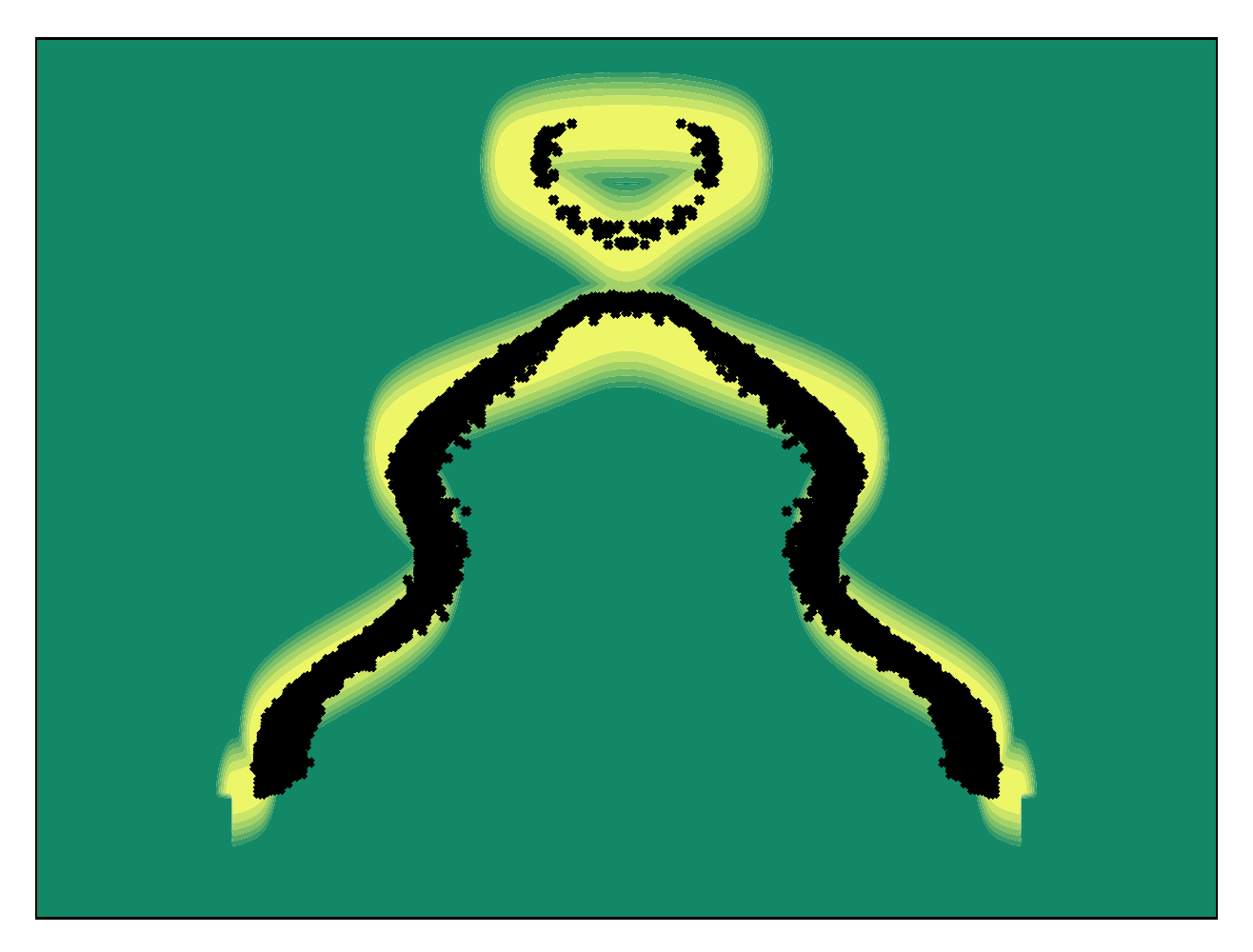}
~
\includegraphics[height=0.125\linewidth]{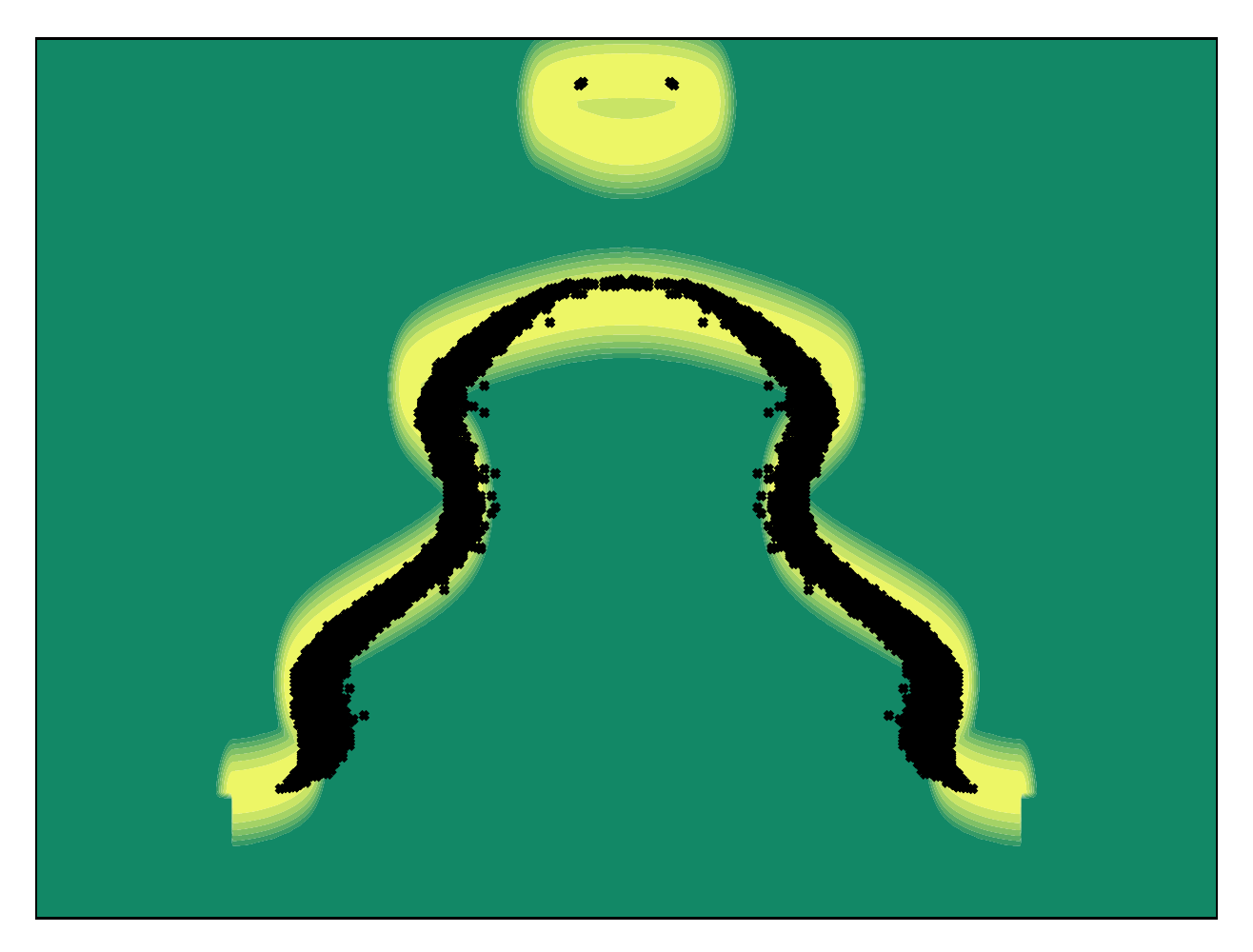}
~
\raisebox{0.005\linewidth}{\includegraphics[height=0.12\linewidth]{colorbar-eps-converted-to.pdf}}

\raisebox{0.0575\linewidth}{300\,Hz}
\quad
\includegraphics[height=0.125\linewidth]{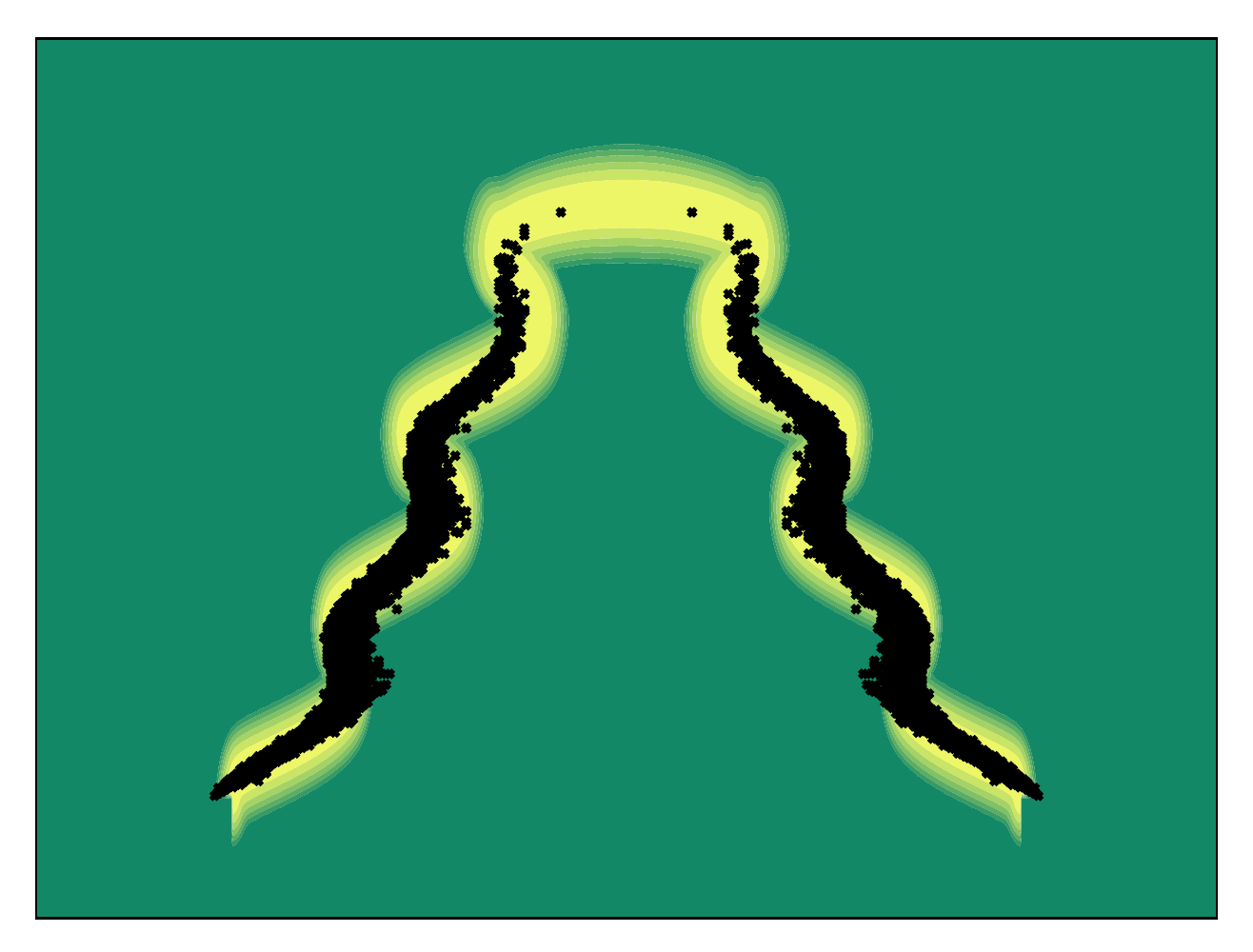}
~
\includegraphics[height=0.125\linewidth]{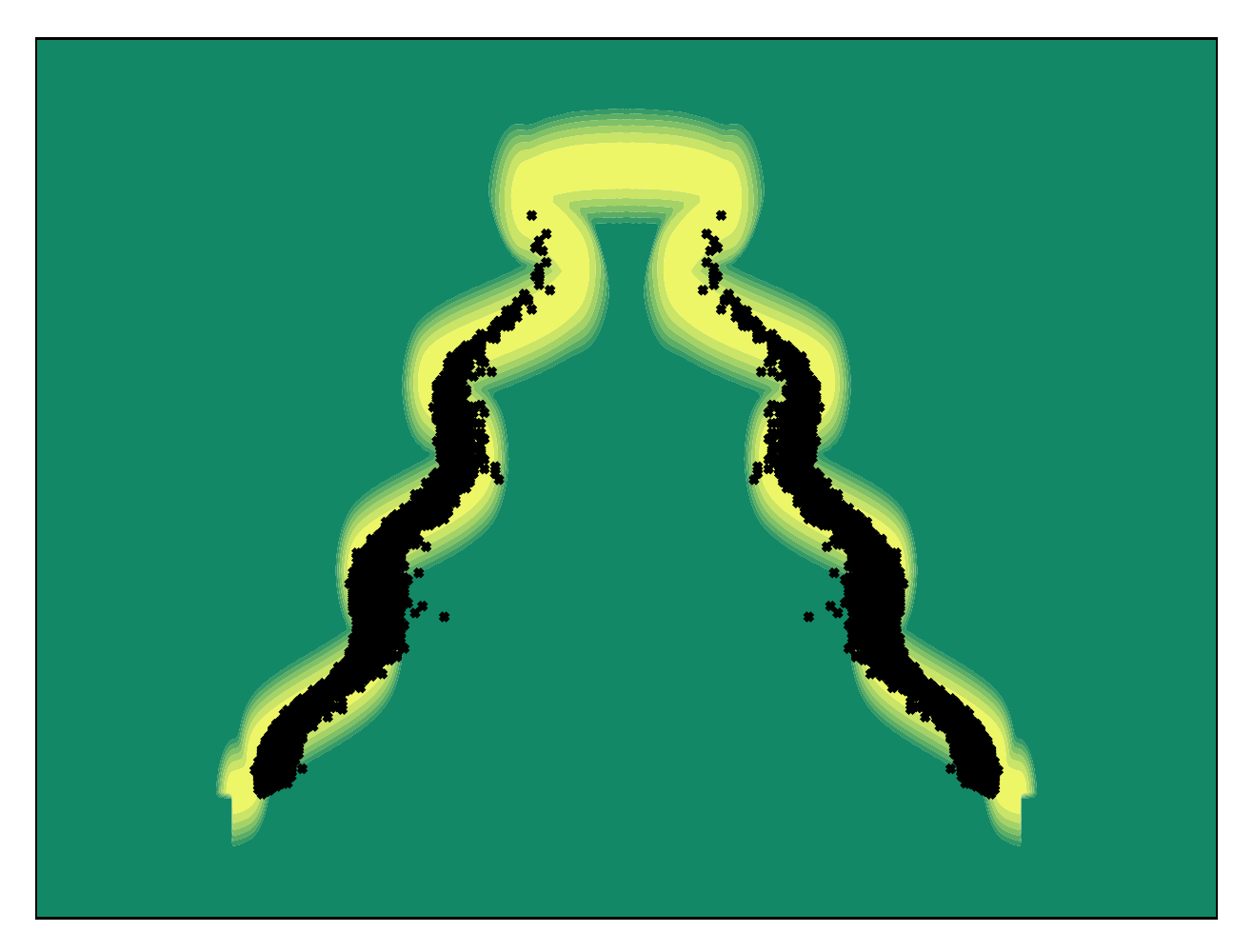}
~
\includegraphics[height=0.125\linewidth]{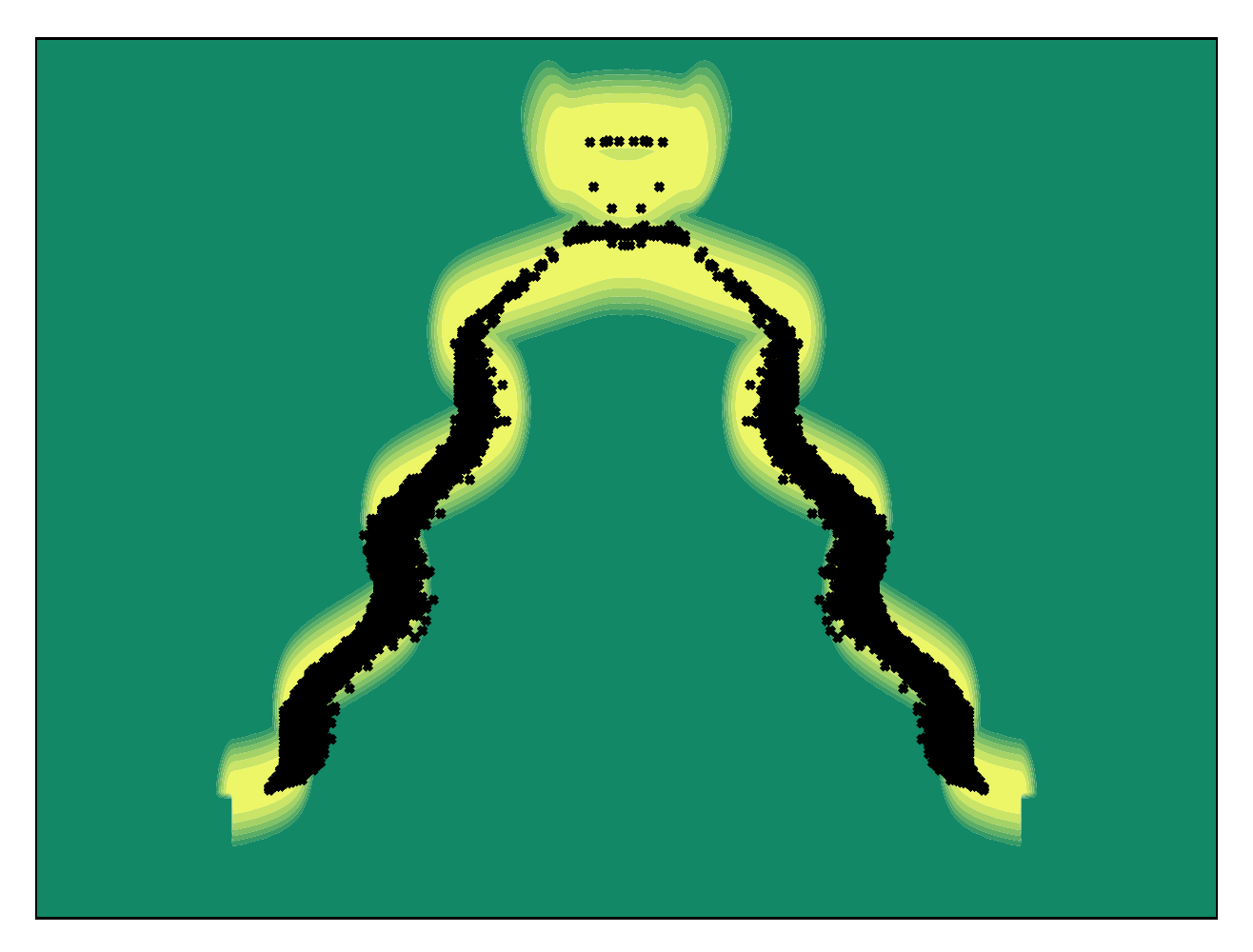}
~
\includegraphics[height=0.125\linewidth]{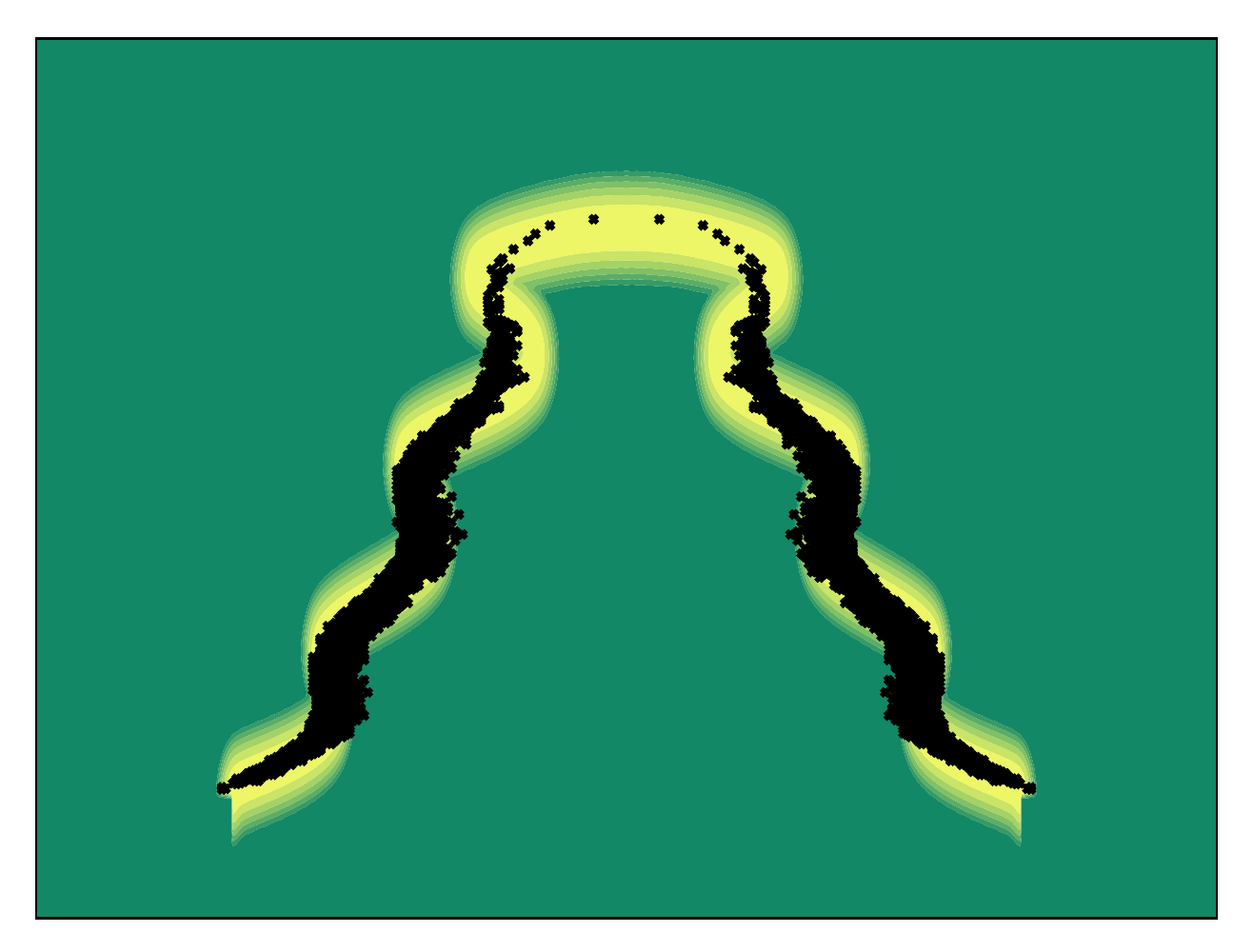}
~
\raisebox{0.005\linewidth}{\includegraphics[height=0.12\linewidth]{colorbar-eps-converted-to.pdf}}

\raisebox{0.0575\linewidth}{400\,Hz}
\quad
\includegraphics[height=0.125\linewidth]{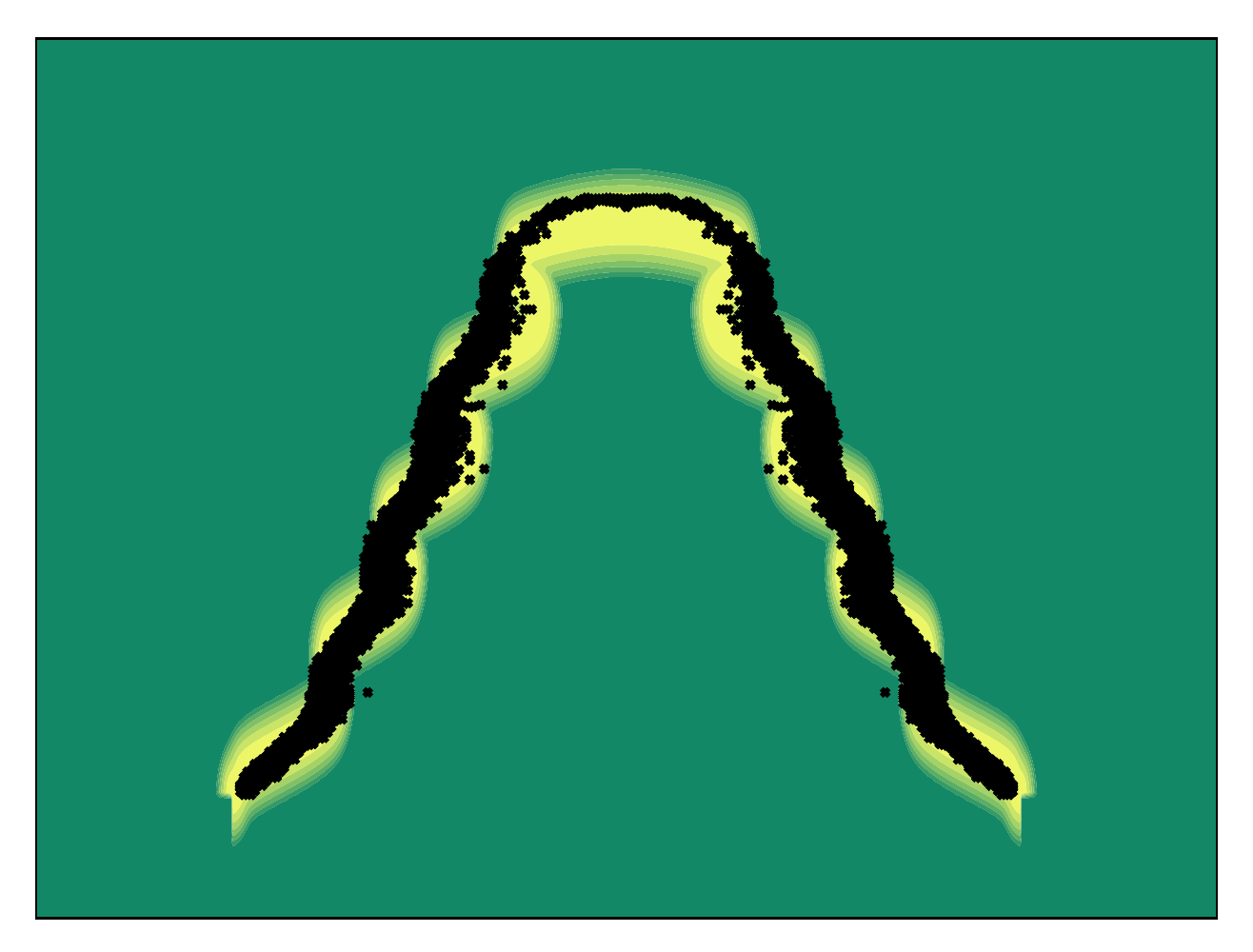}
~
\includegraphics[height=0.125\linewidth]{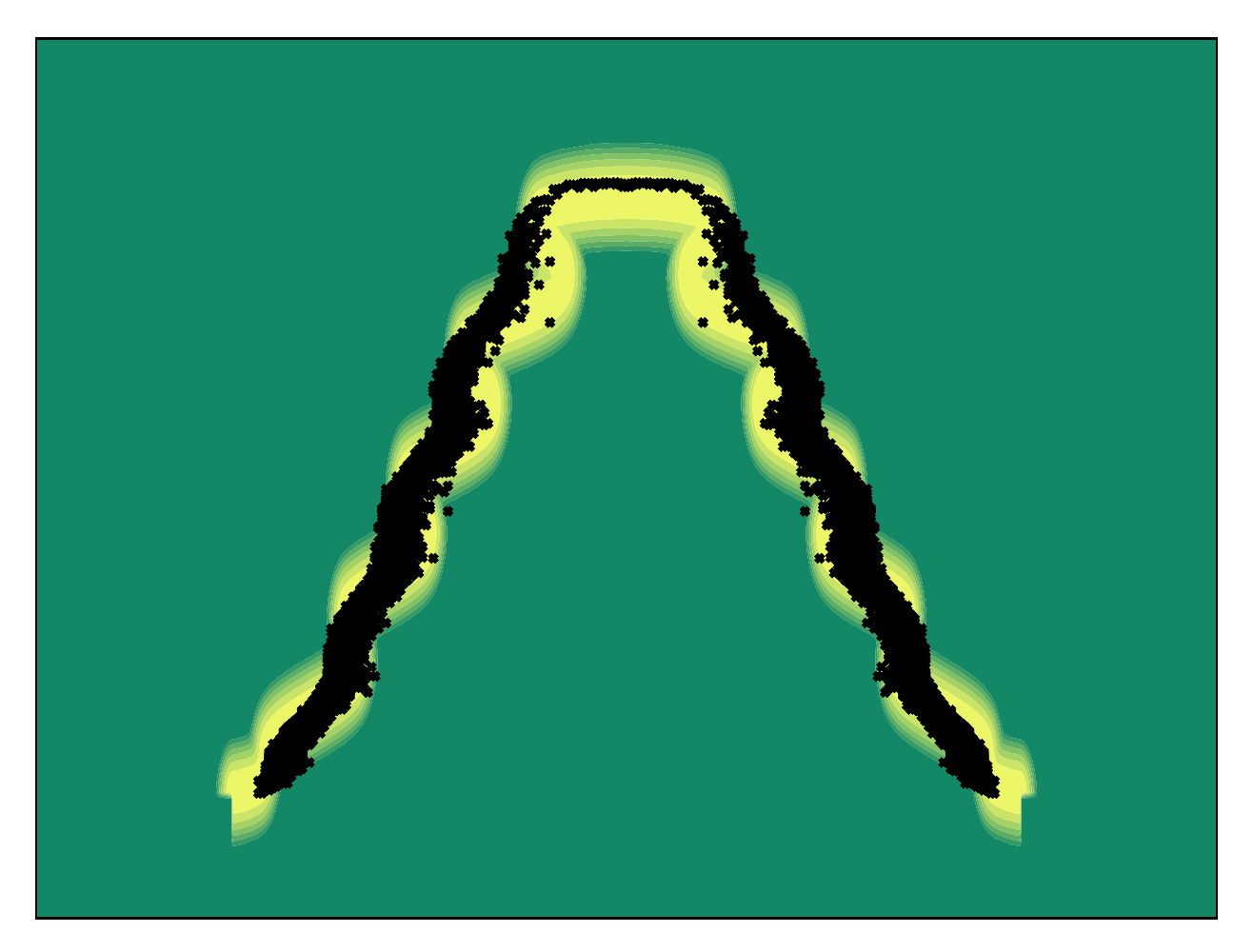}
~
\includegraphics[height=0.125\linewidth]{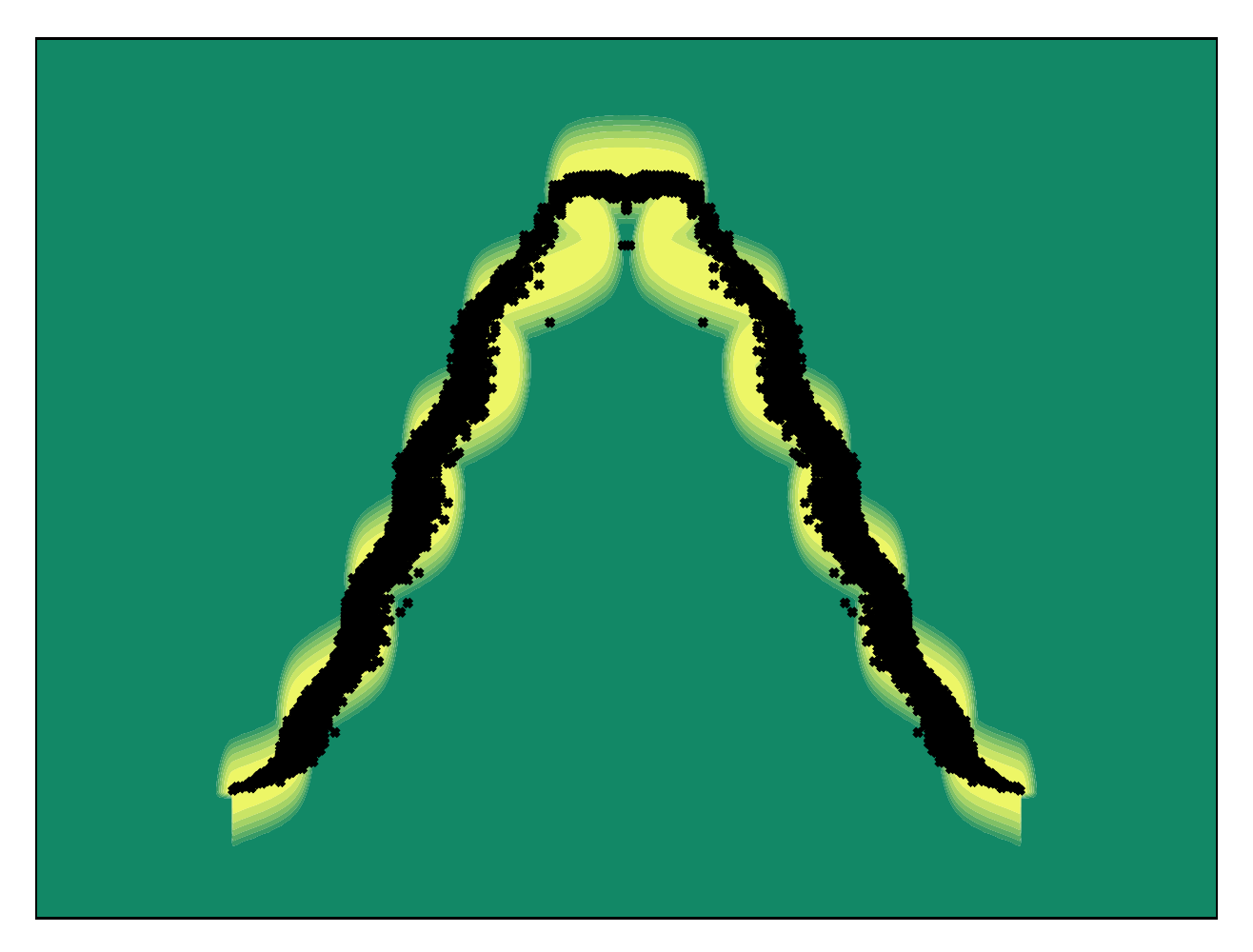}
~
\includegraphics[height=0.125\linewidth]{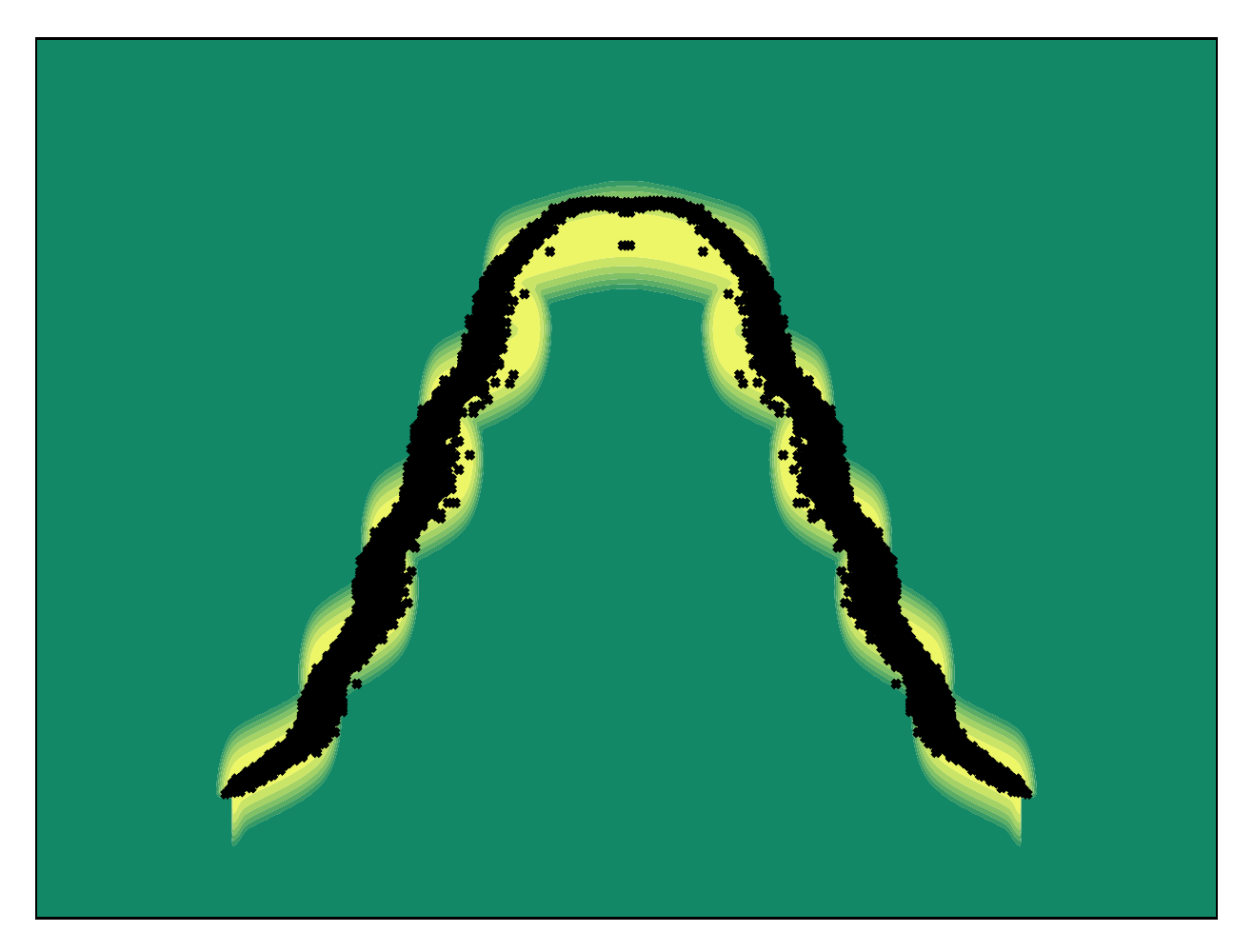}
~
\raisebox{0.005\linewidth}{\includegraphics[height=0.12\linewidth]{colorbar-eps-converted-to.pdf}}
\caption{
Snapshots of log-normalized likelihood (Eq.~\eqref{eq:da:sp:loglik}) over one forcing period after combined state and parameter estimation for 200, 300 and 400\,Hz (top/middle/bottom row, respectively).
Highly likely positions of the flame surface are shown in yellow; less likely positions in green.
The flame surface as detected from experimental images is included (black dots).
}
\label{fig:res:heat}
\end{figure*}

In Fig.~\ref{fig:res:heat}, the outcome of combined state and parameter estimation is visualized for forcing at 200, 300 and 400\,Hz, respectively.
The optimally calibrated reduced-order model accurately captures the perturbations traveling along the flame surface as well as the fuel-air pockets pinching off.
While no individual $G$-equation simulation captures the motion of the flame surface completely, the $G$-equation simulations as an ensemble form an envelope in which the flame surface is fully contained, thus quantifying the uncertainty in the reduced-order model.
While the pinched-off fuel-air pockets are clearly detectable in the experimental images for~200\,Hz, the pinched-off fuel-air pockets are smaller in size for higher frequencies due to the low-pass nature of the flame response, and exist for shorter periods of time.
Although the light intensity is diminished towards the tip of the flame surface, which complicates edge detection and observation in general, the optimally calibrated reduced-order model correctly infers the precise flame dynamics (300\,Hz, left/middle left) that lead to the short-lived fuel-air pocket pinching off~(300\,Hz, middle right).
As the perturbations traveling along the flame surface decrease in magnitude, so does the signal-to-noise ratio.
Combined with the reduced relative frame rate at higher frequencies, the experimental images are ambiguous as to whether a fuel-air pocket pinches off, or the tip of the flame surface only retracts so rapidly because of the high local curvature (400\,Hz, middle right).
This ambiguity is reflected in the ensemble of $G$-equation simulations\revis{, especially in the elevated uncertainty towards the end of the assimilation window}, where some exhibit pinched-off fuel-air pockets with lifespans below the frame rate while others do not \revis{(Fig.~\ref{fig:res:s})}.

\section{Conclusions}
\label{sec:sum}

We develop and test a data-driven reduced-order model of a ducted premixed flame.
This reduced-order model is nonlinear, adaptive and based on physical principles.
\revis{For the first time,} experimental data is rigorously assimilated into the model.
\revis{This is a significant advancement compared to the assimilation of synthetic or simulation data \cite{Yu2019a}.
Firstly, even a direct numerical simulation introduces assumptions into its underlying physical model, e.g.\ regarding the validity of reduced chemical mechanisms, ignored conjugate heat transfer or artificial flame thickening.
Secondly, the experiment produces pinched-off fuel-air pockets unobserved in the available simulation data, which poses a particular challenge to the level-set method.
Thirdly, the experimental data requires more sophisticated data processing in order to identify the flame surface and to remove light emission noise.
}

The model is validated by comparing its behavior with that from experimental data that it cannot observe.
The two key aspects of the analysis are the following:
Firstly, the $G$-equation is a fully nonlinear model.
This includes non-smooth features, e.g.\ cusps on the flame surface, and discontinuities such as topological merging and break-up.
Unlike classical, sensitivity-based approaches designed under linear assumptions, this probabilistic, en\-sem\-ble-based approach successfully models the nonlinearities and delivers an optimally calibrated reduced-order model.
Secondly, the proposed level-set data assimilation framework is based on Bayesian inference~\cite{Yu2019}.
As such, all estimates are equipped with statistically rigorous uncertainty quantification.
This is highly relevant to the design of combustion systems:
In thermoacoustics, for example, slight errors in the model deduced from a single burner could have a large impact on predictions in a different configuration, such as inside an annular combustor~\cite{Juniper2018, Magri2019}.
\revis{
The data assimilation framework was developed for level-set methods in general.
As such, it is readily generalizable to other flame shapes modeled by the $G$-equation.
}
\revis{
In practice, the reduced-order model is optimally calibrated to laboratory experiments, and adapts on-the-fly when new observations become available during operation.
}

This study highlights the role that data can play in the future of combustion modeling for thermoacoustics.
It is increasingly impractical to store data, particularly as experiments become automated~\cite{Rigas2016} and numerical simulations become more detailed.
Rather than store the data itself, the technique in this paper optimally assimilates the data into the parameters of a physics-based model.
With this technique, rapid prototyping of combustion systems can feed into rapid calibration of their reduced-order models and then into gradient-based design optimization.
While it has been shown, e.g.\ in the context of ignition and extinction, that large-eddy simulations become quantitatively predictive when augmented with data~\cite{Labahn2018}, the reduced-order modeling of flame dynamics in turbulent flows remains challenging.
For these challenging situations, this work opens up new possibilities for the development of reduced-order models that adaptively change any time that data from experiments or simulations becomes available.



\section*{Acknowledgments}

The authors would like to thank U.\ Sengupta for the acquisition of the experimental data.
H. Yu is supported by the Cambridge Commonwealth, European \& International Trust under a Schlumberger Cambridge International Scholarship.
L. Magri gratefully acknowledges support from the Royal Academy of Engineering Research Fellowships scheme and the visiting fellowship at the Technical University of Munich -- Institute for Advanced Study, funded by the German Excellence Initiative and the European Union Seventh Framework Programme under grant agreement no. 291763.



\end{document}